\documentclass[11pt,a4paper]{article}
\pdfoutput=1
\usepackage{jcappub}

\usepackage[utf8]{inputenc}
\usepackage[T1]{fontenc}
\usepackage{graphicx,amssymb,amsmath,wasysym}
\usepackage{xcolor}

\newcommand{\mx}{M_X}
\newcommand{\mn}{M_1}

\newcommand{\xpfo}{{z'_\text{FO}}}
\newcommand{\kfo}{\kappa_\text{FO}}
\newcommand{\kt}{\kappa_1}
\newcommand{\gs}{g_\star}
\newcommand{\gsp}{g'_\star}
\newcommand{\gx}{g_X}
\newcommand{\gn}{g_N}
\newcommand{\Td}{T_d}

\begin{document}

\begin{flushright}
FERMILAB-PUB-17-342-T\\
PI/UAN-2017-608FT
\end{flushright}

\title{Hot Leptogenesis\\from Thermal Dark Matter}

\author[1]{Nicolás Bernal}
\emailAdd{nicolas.bernal@uan.edu.co}

\author[2,3]{and Chee Sheng Fong}
\emailAdd{fong@if.usp.br}

\affiliation[1]{Centro de Investigaciones, Universidad Antonio Nariño\\
Carrera 3 Este \# 47A-15, Bogotá, Colombia}

\affiliation[2]{Instituto de Física, Universidade de São Paulo\\
Rua do Matão 1371, 05508-090 São Paulo, Brazil}

\affiliation[3]{Theoretical Physics Department, Fermi National Accelerator Laboratory\\
PO Box 500, Batavia, IL 60510, U.S.A}

\abstract{
In this work, we investigate a scenario in which heavy Majorana Right-Handed Neutrinos (RHNs) 
are in thermal equilibrium with a dark sector with temperature higher than the 
Standard Model (SM) thermal bath. Specifically, we consider the scenario in which thermal 
Dark Matter (DM) abundance is fixed from the freeze-out of DM annihilations into RHNs. 
Due to the inert nature of the RHNs, we show that it is possible for the two sectors to remain 
thermally decoupled by having more than two generations of the RHNs. 
The hotter temperature implies higher abundances of DM and RHNs with the following consequences.
For leptogenesis, an enhancement in efficiency up to a factor of 51.6 can be obtained, 
though a resonant enhancement of CP violation is still required due to an upper mass bound
of about 4~TeV for the RHNs.
For the DM, an enhanced annihilation cross section up to a factor of 51.6 is required to
obtain the correct DM abundance. 
This scenario can be probed via indirect detection of DM annihilating 
into RHNs, which then decay into $h\,\nu$, $Z\,\nu$ and $W^{\pm} \ell^{\mp}$ 
with an enhanced annihilation cross section above the typical thermal value.

}

\maketitle


\section{Introduction}

The Baryon Asymmetry of the Universe (BAU), Dark Matter (DM) and nonzero neutrino masses 
represent strong evidences of physics beyond the Standard Model (SM) to date.
In principle the DM and the BAU could be unrelated.
They have indeed often been approached separately in the literature.
Nevertheless one could entertain the idea that they have a common origin, 
motivated by the fact that the ratio of the abundances of dark and baryonic matter 
$\Omega_\text{DM}/\Omega_\text{B}\sim 5$ is of the same order.
This would suggest a common mechanism for the origin of the two species.
Such a possibility is in the framework of Asymmetric Dark Matter 
(ADM)~\cite{Nussinov:1985xr,Roulet:1988wx,Barr:1990ca}, 
where one speculates that in the dark sector there is a matter anti-matter asymmetry 
that is related to the one in the visible sector, the BAU.
For instance, in ref.~\cite{Bernal:2016gfn}, a model-independent approach was taken 
to study scenarios in which DM asymmetry and SM matter anti-matter asymmetry are shared 
through effective operators. A number of other possibilities, this time assuming a 
symmetric weakly interacting massive particle (WIMP) have also been considered e.g. 
`dark matter assimilation'~\cite{DEramo:2011dhr}, `baryomorphosis'~\cite{McDonald:2011zza,McDonald:2011sv}, 
`WIMPy baryogenesis'~\cite{Cui:2011ab, Bernal:2012gv, Bernal:2012hia, Bernal:2013bga, Bernal:2013gsa, Racker:2014uga}.

Type-I seesaw~\cite{Minkowski:1977sc,Yanagida:1979as,
GellMann:1980vs,Glashow:1979nm,Schechter:1980gr,Mohapatra:1980yp}
has always been appreciated for its simplicity and elegance in explaining
the small neutrino masses and BAU through leptogenesis~\cite{Fukugita:1986hr,Davidson:2008bu,Fong:2013wr}. 
It involves a new scale in which lepton number $L$ is violated and new degrees of freedom 
i.e. the RHNs are at play. Light neutrino masses are explained by having a large 
lepton-number-violating scale aptly known as the seesaw scale $\Lambda_{{\rm seesaw}}$.
While neutrino masses do not impose a lower bound on $\Lambda_{{\rm seesaw}}$,
they impose an upper bound of $\Lambda_{{\rm seesaw}}\lesssim10^{16}$
GeV to keep the neutrino Yukawa couplings perturbative $\lambda\lesssim4\pi$.
Regarding leptogenesis, it imposes both upper and lower bounds
on $\Lambda_{{\rm seesaw}}$. The upper bound $\Lambda_{{\rm seesaw}}\lesssim 10^{14}$
GeV is to keep under control the washout of lepton asymmetry from
$\Delta L=2$ scatterings with $\lambda\lesssim 1$. 
The lower bound is model dependent. Assuming hierarchical masses of the RHNs, 
the lower bound of $\Lambda_{{\rm seesaw}}\gtrsim 10^{9}$~GeV is obtained to have sufficient
CP violation for leptogenesis (due to the measured light neutrino
mass differences) \cite{Davidson:2002qv}. By allowing almost degenerate masses of RHNs, 
one can realize resonant leptogenesis~\cite{Pilaftsis:1997jf} which allows to lower 
the seesaw scale down to the scale of temperature before electroweak sphalerons freeze-out at 
$T_{\rm EWsp} = 132$ GeV~\cite{DOnofrio:2014rug} such that the baryon asymmetry 
can be induced through leptogenesis. There is another interesting and physically different 
mechanism of leptogenesis through RHN oscillation first proposed by
Akhmedov, Rubakov and Smirnov (ARS)~\cite{Akhmedov:1998qx}. In ARS leptogenesis, 
the seesaw scale is required to be around GeV and the oscillation of the RHNs 
commences at around $T \lesssim 10^6$~GeV. 

In this work we aim to connect DM together with type-I seesaw thereby 
establish the connection between DM, BAU and neutrino masses. 
In particular, we explore the possibility that the dark sector, i.e. 
where DM resides, starts up \emph{hotter} than the SM sector after inflation. 
The hotness of this sector can be attributed to its stronger
coupling to inflaton compared to the SM sector~\cite{Hodges:1993yb,Berezhiani:1995am,Dev:2013yza,Kane:2015qea}.
Alternatively, the difference of temperatures could have been dynamically generated 
through cannibalization within the dark sector~\cite{Carlson:1992fn, Hochberg:2014dra, Hochberg:2014kqa, Bernal:2015bla, Bernal:2015lbl, Lee:2015gsa, Choi:2015bya,Hansen:2015yaa, Bernal:2015ova, Bernal:2015xba, Kuflik:2015isi, Hochberg:2015vrg, Choi:2016hid, Pappadopulo:2016pkp, Heikinheimo:2016yds, Farina:2016llk, Choi:2016tkj, Bernal:2017mqb, Choi:2017mkk, Heikinheimo:2017ofk, Ho:2017fte, Kuflik:2017iqs}.
We will not discuss about a particular realization of hotter dark sector but take it as an initial condition.

We further assume that the RHNs are in thermal equilibrium with the dark
sector such that they are hotter as well. This is achieved in our setting 
by having DM particles which annihilate to RHNs and the final DM abundance 
is determined from the freeze-out when the annihilation becomes inefficient. 
In this work we will not consider the possible impacts of the asymmetries in the DM. 
To be as model independent as possible,\footnote{We note that UV-complete models where the DM annihilates into RHNs, in dark sectors that can be hotter than the SM, have been considered, see e.g. refs~\cite{Escudero:2016tzx, Escudero:2016ksa}.}
we take the annihilation cross section 
as free parameter bounded only by unitarity~\cite{Griest:1989wd,Hui:2001wy}. 

The requirement to maintain a hotter dark sector and the RHNs compared
to the SM sector gives rise to several interesting results which we
will now highlight:
\begin{enumerate}
\item[(1)] From the observed light neutrino mass differences, more than two generations
of RHNs are needed due to the requirement that at least
one RHN responsible for leptogenesis should not thermalize with
the SM sector. For the scenario with \emph{three} RHNs, 
this places an upper bound on the mass of the lightest light neutrino which becomes stronger 
as RHNs become hotter. 

\item[(2)] With hotter RHNs, their thermal abundances with respect to the SM particles 
are enhanced, as the lepton asymmetry generated from their decays. 
The decoupled nature of decaying RHNs implies that they will decay very out-of-equilibrium where 
washout processes, lepton flavor effects~\cite{Barbieri:1999ma,Abada:2006fw,Nardi:2006fx,Abada:2006ea}
and thermal effects~\cite{Covi:1997dr,Giudice:2003jh, Biondini:2015gyw, Biondini:2016arl} are negligible. 

\item[(3)] With hotter DM, due to the enhancement in its abundance, 
the annihilation cross section required to obtain the correct
DM abundance will be proportionally higher than the standard thermal cross section of 
few $10^{-26}\,{\rm cm}^3{\rm /s}$~\cite{Steigman:2012nb}.

\item[(4)] With the late decays of hotter RHNs, they can dominate the energy density of the Universe 
and upon decay, inject significant entropy into the SM sector. This dilution will partially compensate the gain 
in abundance from hotness. Denoting the \emph{efficiency} schematically as $\eta = {\rm hotness}/{\rm dilution}$, 
we find an upper bound of $\eta < 51.6$.\footnote{The precise definition of $\eta$ is in eq.~\eqref{eq:efficiency}.}
This implies the overall gain by the same factor in the efficiency of leptogenesis as discussed 
in (2) as well as the enhancement of annihilation cross section as discussed in (3).

\end{enumerate}

The paper is organized as follows. In section~\ref{sec:HL}, 
we will discuss the required conditions to achieve \emph{hot} leptogenesis i.e. 
leptogenesis which proceeds through decays of RHNs with thermal abundance at 
a temperature higher than the SM's one.
The production mechanism for the DM relic abundance is presented in section~\ref{sec:DM}.
Section~\ref{ID} is devoted to the study of the indirect detection prospects, in particular from gamma-rays.
Finally, the conclusions are presented in section~\ref{sec:conclu}.

\section{Hot Leptogenesis}
\label{sec:HL}

In the section, we first review the type-I seesaw mechanism and 
fix the notation. In section~\ref{sec:HL_conditions}, we will discuss 
the basic requirement to achieve \emph{hot} leptogenesis. In section~\ref{sec:connection_DM}, 
we will establish the connection between the RHNs and DM and spell out the three possible 
scenarios we will consider in this work. In section~\ref{sec:scatt_osc}, 
we will discuss the coherence and the oscillations of RHNs, and the conditions required to avoid 
thermalization between the dark and the visible sectors. Finally, in section~\ref{sec:realizations_leptogenesis}, 
we will discuss different possible realizations of hot leptogenesis. In particular,
we will derive a new lower mass bound on the lightest RHN under the assumption of hierarchical RHNs. 

The type-I seesaw Lagrangian is given by 
\begin{equation}
-{\cal L} \supset \frac{1}{2} M_{i}\,\overline{N}_{i}^{c}N_{i}
+\lambda_{\alpha i}\,\overline{\ell}_{\alpha}\,\epsilon\,\phi^{*}\,N_{i}+{\rm H.c.}\,,
\end{equation}
where we have chosen a basis where the Majorana mass term of RHNs $N_{i}$
$\left(i=1,\,2,\,...,\,n\right)$ is diagonal and the charged lepton Yukawa
(not shown) is also diagonal, while $\ell_{\alpha}$ $\left(\alpha=e,\mu,\tau\right)$
and $\phi$ are respectively the SM lepton and Higgs doublets and
$\epsilon$ is the totally antisymmetric tensor of $SU(2)$. 

The neutrino Yukawa is conveniently parametrized following Casas and
Ibarra as~\cite{Casas:2001sr}
\begin{equation}
\lambda_{\alpha i} = \frac{1}{v}\,\left(U^{*}\sqrt{\hat{m}}\,R\,\sqrt{\hat{M}}\right)_{\alpha i}\,,\label{eq:Casas_Ibarra_Param}
\end{equation}
where $v\equiv\left\langle \phi\right\rangle\simeq 174$~GeV is the Higgs
vacuum expectation value, $U$ is the Pontecorvo-Maki-Nakagawa-Sakata
matrix \cite{Pontecorvo:1957qd,Maki:1962mu}, the light neutrino masses
$\hat{m}={\rm diag}\left(m_{1},m_{2},m_{3}\right)$, the RHN 
masses $\hat{M}={\rm diag}\left(M_{1},M_{2},...,M_{n}\right)$
and $R$ a complex $3\times n$ orthogonal matrix. Without loss of
generality, we will fix $M_{1}<M_{2}<...<M_{n}$. For our purpose,
we will just use the global best fit values of ref.~\cite{Esteban:2016qun}:
$\Delta m_{{\rm sol}}^{2}\equiv m_{2}^{2}-m_{1}^{2}=7.5\times10^{-5}\,{\rm eV}^{2}$
and $\Delta m_{{\rm atm}}^{2}\equiv\left|m_{3}^{2}-m_{1}^{2}\right|=2.5\times10^{-3}\,{\rm eV}^{2}$.

For $n=2$, $RR^{T}={\rm diag}\left(0,1,1\right)$ and $R^{T}R=I_{2\times2}$.
For $n=3$, $RR^{T}=R^{T}R=I_{3\times3}$. For
$n>3$, $RR^{T}=I_{3\times3}$ while $R^{T}R$ is not fixed in general.
The parametrization in eq.~\eqref{eq:Casas_Ibarra_Param} and the forms of
$R$ make sure that we obtain the correct neutrino mass spectrum and
mixing with the light neutrino mass matrix given by
\begin{equation}
m_{\nu} = v^{2}\,\lambda\,\hat{M}^{-1}\,\lambda^{T}
\label{eq:typeI_nu_mass}
\end{equation}
at the leading order, assuming $\lambda_{\alpha i}\,v\ll M_{i}$.

\subsection{Conditions for Hot Leptogenesis}
\label{sec:HL_conditions}

Next we would like to investigate the conditions under which $N_i$ 
can remain decoupled from the thermal bath. We will do this by comparing 
the $N_i$ decay rate $\Gamma_i$ to the expansion Hubble rate $H$ at $T'=M_i$, where $T'$ is the 
dark sector temperature. If $\Gamma_i < H$ at $T'=M_i$, 
$N_i$ is out of thermal equilibrium with the SM sector 
at $T'\geq M_i$ through decays, inverse decays and scatterings.
On the other hand, if $\Gamma_i > H$ at $T'=M_i$, 
the decays and inverse decays (and likely also scatterings) equilibrate the two sectors 
and hot leptogenesis is not viable.\footnote{The scattering 
rate of $N_i$ with the SM particles at $T > M_i$ 
goes like $\sim 0.1 \frac{T}{M_i} \Gamma_i$~\cite{Cline:1993bd} 
and hence is always slower than the Hubble expansion rate if $\Gamma_i < H$ at $T'=M_i$. 
We will discuss further the impacts of these processes in section \ref{sec:scatt_osc}.} 
The decay width of $N_{i}$ is given by
\begin{equation}
\Gamma_{i} = \frac{\left(\lambda^{\dagger}\lambda\right)_{ii}M_{i}}{8\pi}\equiv\frac{\tilde{m}_{i}\,M_{i}^{2}}{8\pi\,v^{2}}\,,
\end{equation}
where we defined the \emph{effective} neutrino masses as
\begin{equation}
\tilde{m}_{i} \equiv \frac{\left(\lambda^{\dagger}\lambda\right)_{ii}\,v^{2}}{M_{i}}
=\left(R^{\dagger}\,\hat{m}\,R\right)_{ii}\,.
\label{eq:meff}
\end{equation}
In the second equality above, we have made use of the parametrization
of eq.~\eqref{eq:Casas_Ibarra_Param}. Additionally, the Hubble rate
is given by
\begin{equation}
H  =  \sqrt{\frac{8\pi}{3M_{{\rm Pl}}^{2}}\,\rho}\,,
\label{eq:Hubble}
\end{equation}
where $M_{{\rm Pl}}=1.22\times10^{19}$ GeV is the Planck mass and
the total energy density of the Universe $\rho$ is
\begin{eqnarray}
\rho & = & \frac{\pi^{2}}{30}\,\gsp\,T'^{4}+\frac{\pi^{2}}{30}\,\gs\,T^{4} \nonumber \\
& = & \frac{\pi^{2}}{30}\left(\gsp\,\kappa^{4}+ \gs\right)T^{4}  \nonumber \\
& = & \frac{\pi^{2}}{30}\left(\gsp+\frac{\gs}{\kappa^{4}}\right)T'^{4}\,,
\label{eq:energy_density_universe}
\end{eqnarray}
with $T$ and $T'$ being the temperatures of the visible and dark sectors, respectively. 
Additionally, $\kappa\equiv T'/T$ and $\gsp$ corresponds to 
the relativistic degrees of freedom in the dark sector including the RHNs. 
Anticipating the next sections, from the last line 
of eq.~\eqref{eq:energy_density_universe},
we see that at a given $T'$, the Hubble rate is `slower' for $\kappa > 1$ 
and $\gsp < \gs$.\footnote{With respect to the visible or SM temperature $T$, 
from the second line of eq.~\eqref{eq:energy_density_universe}, it is clear 
that the Hubble rate is \emph{always} faster with additional contribution from the dark sector.
Nevertheless since we are interested in the annihilation rate of DM 
with temperature $T'$, we should compare it with the Hubble rate at $T'$ 
given by the last line of eq.~\eqref{eq:energy_density_universe} where the
effective relativisitc degrees of freedom decreases with $\kappa$.}
For thermal freeze-out of DM, this implies 
that the DM has to be heavier and/or the annihilation cross section has to be smaller. 
Nevertheless, this is partially compensated by having a larger abundance 
of DM due to $\kappa > 1$.

Comparing $\Gamma_i$ and $H$ at $T'=M_{i}$, we have
\begin{eqnarray}
K_{i}\equiv\left.\frac{\Gamma_{i}}{H}\right|_{T'=M_{i}} 
& = & \frac{\tilde{m}_{i}\,M_{{\rm Pl}}}{16\pi v^{2}\sqrt{\frac{\pi^{3}}{45}\left(\gsp+\frac{\gs}{\kappa^{4}}\right) }}\nonumber \\
& = & 9.66 \times10^{2}\left(\frac{\tilde{m}_{i}}{0.1\,{\rm eV}}\right)
\frac{1}{\sqrt{\gsp+\frac{\gs}{\kappa^{4}}}}\,.
\label{eq:K_decay}
\end{eqnarray}
For instance, taking $\gs=106.75$,
$\gsp=6$ (e.g. three relativistic RHNs) and
$\kappa=1$, one gets
\begin{equation}
K_{i} = 91.0\left(\frac{\tilde{m}_{i}}{0.1\,{\rm eV}}\right)\,.
\end{equation}

Notice that we have $\tilde{m}_{i}=m_{1}\left|R_{1i}\right|^{2}+m_{2}\left|R_{2i}\right|^{2}+m_{3}\left|R_{3i}\right|^{2}$.
For $n=2$, the lightest neutrino is massless and hence $\tilde{m}_{i}\geq\sqrt{\Delta m_{{\rm sol}}^{2}}$
which gives  $K_{i=1,\,2}\geq 7.88$. 
Hence we conclude that in order to
have the possibility of hot leptogenesis ($\kappa>1$), we require \emph{at
least} $n>2$.\footnote{
For ARS leptogenesis which occurs due to oscillations of RHN,
it is possible that the SM and dark sectors do not thermalize even with $n=2$. 
See the discussion in section~\ref{sec:realizations_leptogenesis} for details.
} 
For $n=3$, we have $\tilde{m}_{i}\geq m_{l}$ where
$m_{l}$ is the mass of the lightest light neutrino with $l=1$ $(3)$
for normal (inverted) mass ordering of light neutrino masses. For $n>3$, there is in principle
no lower bound on $\tilde{m}_{i}$ and hot leptogenesis is always possible.

For definiteness in this work, we stick to the most interesting case $n=3$.
In this case, the \emph{absolute scale} of neutrino will
determine whether hot leptogenesis is possible. 
Due to orthogonality conditions $R^{T}R=RR^{T}=I_{3\times3}$, 
if we assume $N_{1}$ to be the one which couples weakly 
to the SM sector $K_1 < 1$, due to the measured mass differences of neutrinos, 
$N_{2}$ and $N_{3}$ will couple strongly to the SM sector with $K_{2}$, $K_{3} > 1$.
This can be understood as follows.
For Normal mass Ordering (NO) with $\tilde{m}_{1}=m_{1}$, orthogonality
conditions imply 
\begin{eqnarray}
\tilde{m}_{1} & = & m_{1},\\
\tilde{m}_{2} & = & m_{2}\left|1-R_{32}^2\right|+m_{3}\left|R_{32}\right|^{2},\\
\tilde{m}_{3} & = & m_{2}\left|R_{32}\right|^{2}+m_{3}\left|1-R_{32}^{2}\right|.
\end{eqnarray}
In this case, we have $\min\left(\tilde{m}_{2},\tilde{m}_{3}\right)=m_{2}\geq\sqrt{\Delta m_{{\rm sol}}^{2}}$.
For Inverse mass Ordering (IO) with $\tilde{m}_{1}=m_{3}$, orthogonality
conditions imply 
\begin{eqnarray}
\tilde{m}_{1} & = & m_{3},\\
\tilde{m}_{2} & = & m_{1}\left|1-R_{22}^{2}\right|+m_{2}\left|R_{22}\right|^{2},\\
\tilde{m}_{3} & = & m_{1}\left|R_{22}\right|^{2}+m_{2}\left|1-R_{22}^{2}\right|.
\end{eqnarray}
This gives us $\min\left(\tilde{m}_{2},\tilde{m}_{3}\right)=m_{1}\geq\sqrt{\Delta m_{{\rm atm}}^{2}}$.
In either cases, we have $K_{2,\,3}\gg 1$.
In the rest of the work, we will assume that $N_{1}$ is weakly
coupled to SM with $K_{1}<1$ while $N_{2}$ and $N_{3}$ are strongly
coupled with $K_{2,\,3}\gg1$. 

\subsection{Connection with the Dark Sector}
\label{sec:connection_DM}

To establish a connection with DM, we assume that the dark sector
contains a DM $X$ with mass $M_{X}$ which can annihilate
to RHNs through $X\,X\to N_i\,N_i$ and the final abundance
of $X$ is determined by the freeze-out of the previous reaction at $T'_{\rm FO}$.
We denote the ratio between the dark and visible temperatures at freeze-out as
\begin{equation}
\kfo \equiv \frac{T'_{\rm FO}}{T_{\rm FO}}\,.
\label{eq:kappa_FO}
\end{equation}

We will consider the case where the annihilation cross section is $s$-wave dominated
and bounded only by unitarity~\cite{Griest:1989wd,Hui:2001wy}. 
Furthermore, we will define the \emph{abundance} of
the species $a$ by normalizing its number density with the SM entropic
density $s=\frac{2\pi^{2}}{45}\,\gs\,T^{3}$
as follows $Y_a \equiv \frac{n_a}{s}$.
In order not to suppress the abundance of $N_1$, which is responsible for leptogenesis, 
we assume that $N_1$ is relativistic at freeze-out, i.e.
\begin{equation}
M_1 < T'_{\rm FO} = \frac{M_X}{z'_{\rm FO}},
\label{eq:N1_relativistic}
\end{equation}
where we have defined $z'_{\rm FO} \equiv \frac{M_X}{T'_{\rm FO}}$.
Now we will further define the temperature of $N_1$ with respect to that 
of the SM as
\begin{equation}
\kappa_1 \equiv \frac{T'_{N_1}}{T}.
\label{eq:kappa_1}
\end{equation}
In principle, $\kappa_{\rm FO}$ can be different from $\kappa_1$ because 
the decays of $N_2$ and $N_3$ heat up the SM thermal bath resulting in 
$\kappa_1 \leq \kappa_{\rm FO}$, as we will discuss later.

With the restriction eq.~\eqref{eq:N1_relativistic}, 
the abundance of $N_1$ after the DM freeze-out, 
but before their decays is given by
\begin{equation}
Y_{N_1} = \frac{135\,\zeta(3)}{4\pi^{4}\,\gs}\kappa_1^{3}\,.
\label{eq:N1_abundance}
\end{equation}
The abundance is conserved as long as there is no
entropy injection to the SM sector i.e. $\kappa_1$ remains constant.
One crucial point is in order. Since $N_1$ has an abundance which is \emph{conserved} 
(up to possible dilution from which will take into account) 
and it decays \emph{out-of-equilibrium}, it can dominate the energy density of the Universe 
and hence its contribution to the Hubble rate through the total energy density,
eq.~\eqref{eq:energy_density_universe}. The out-of-equilibrium condition implies the decaying 
temperature of $N_1$ to be $T'_d < M_1$, and the corresponding temperature in the SM 
to be
\begin{equation}\label{eq:outofeq}
T_d < \frac{M_1}{\kappa_1}\,.
\end{equation}
The above requirement also makes sure that the inverse decay 
and temperature equilibration through 2-to-2 scatterings are not effective.
Before its decay, $N_1$ contributes to the energy density 
of the Universe is as follows
\begin{eqnarray}
\rho & = & M_1 \,s\, Y_{N_1}+\frac{\pi^{2}}{30}\,\gs\,T^{4}\nonumber \\
 & = & M_1 \frac{3 \zeta(3)}{2\pi^2}\,\kappa_1^3\, T^3+\frac{\pi^{2}}{30}\,\gs\,T^{4}\,,
\end{eqnarray}
where we have assumed that there is 
no other relativistic degrees of freedom remaining in the dark sector.

The $N_1$ decaying temperature $T'_d$ or $T_d$ can be solved by setting 
its decay rate equal to the Hubble rate $\Gamma_1 = H$ and we obtain:
\begin{equation}
\frac{\pi^2}{30}\gs \,T_d^4 + M_1 \frac{3 \zeta(3)}{2\pi^2}\kappa_1^3 \,T_d^3
= \frac{3\, \tilde m_1^2 \,M_1^4 \,M_{\rm Pl}^2}{(8\pi)^3 \,v^4}\,.
\label{eq:decaying_temperature}
\end{equation}
From the above, we can solve $\tilde m_1$ in term of $T_d$ and $\kappa_1$. 
While the upper bound on $T_d$ is given by eq.~\eqref{eq:outofeq}, 
the lower bound is to have $N_1$ decays before electroweak sphalerons 
freeze-out at $T_{\rm EWsp}$ such that a lepton asymmetry can also induce a baryon asymmetry:
\begin{equation}
T_d > T_{\rm EWsp}.
\label{eq:decaying_temperature_lower_bound}
\end{equation}
The lower bound is actually less strict since the SM 
thermal bath can be reheated from $N_1$ decays and electroweak sphalerons 
can be `reactivated' to convert the lepton asymmetry to baryon asymmetry. 
For leptogenesis, the treatment of this situation needs modification since 
it can take place after electroweak symmetry breaking.
Taking into account this possibility, the lower bound becomes
\begin{equation}
T_d > \frac{T_{\rm EWsp}}{\kappa_1} \equiv T_d^*.
\label{eq:reactivated_bound}
\end{equation}
We will consider this possibility as well in our study.
While hot leptogenesis is restricted to $\kappa_1 > 1$, 
eqs.~\eqref{eq:outofeq} and \eqref{eq:decaying_temperature_lower_bound} also implies 
an upper bound on $\kappa_1$
\begin{equation}
\kappa_1 < \frac{M_1}{T_{\rm EWsp}}\,.
\label{eq:kappa1_bound}
\end{equation}
Considering the possibility that $N_1$ decays reheat the SM thermal bath 
beyond $T_{\rm EWsp}$ (eq.~\eqref{eq:reactivated_bound}), 
there is in principle no upper bound on $\kappa_1$.

Substituting eq.~\eqref{eq:outofeq} 
into eq.~\eqref{eq:decaying_temperature}, we obtain an upper bound on $\tilde m_1$ 
\begin{equation}
\tilde m_1 < \sqrt{\frac{45 \zeta (3)}{\pi^4} + \frac{\gs}{\kappa_1^4}}
\frac{16 \pi^{5/2}}{3 \sqrt{5}} \frac{v^2}{M_{\rm Pl}}\,.
\end{equation}
The bound above in turn implies a bound on the lightest light neutrino
\begin{equation}
m_l \leq \tilde m_1 < 1.035 \times 10^{-4} 
\sqrt{\frac{45 \zeta (3)}{\pi^4} + \frac{\gs}{\kappa_1^4}}\,{\rm eV},
\label{eq:ml_bound}
\end{equation}
which represents a necessary condition for hot leptogenesis with $n = 3$.
For $\kappa_1 \gg 1$, we obtain the strongest bound of 
$m_l < 7.7 \times 10^{-5}$ eV.\footnote{By $\kappa_1 \gg 1$, we are 
studying the limiting case with $\kfo \to \infty$. 
}
This also implies that if we were to determine $m_l$ experimentally, 
we can put an upper bound on $\kappa_1$ through eq.~\eqref{eq:ml_bound}.

$N_1$ plays important roles in both leptogenesis and DM. 
Its hotness with respect to the SM sector gives a boost factor of $\kappa_1^3$ 
to leptogenesis as in eq.~\eqref{eq:N1_abundance}. 
On the other hand, its late decay and entropy injection to the SM sector 
dilutes both the abundance of $B-L$ (baryon minus lepton number) asymmetry and DM. 
We can estimate the dilution from $N_1$ decay as follows. 
In the sudden decay approximation of $N_1$, the conservation of the energy density at $T'=T'_d$ implies
\begin{equation}\label{eq:Ttilde}
M_1 \frac{3 \zeta(3)}{2\pi^2} \,\kappa_1^3\, T_d^3 + \frac{\pi^2}{30} \,\gs \,T_d^4 
= \frac{\pi^2}{30} \,\gs \,\tilde T^4\,,
\end{equation}
where $\tilde T$ is the SM temperature after $N_1$ decays. 
Solving for $\tilde T$, we obtain
\begin{equation}
\tilde T = T_d \left(\frac{45 \zeta(3)\, \kappa_1^3}{\pi^4\, \gs} \frac{M_1}{T_d} + 1 \right)^{1/4}\,.
\end{equation}
Dilution $d$ due to the decay of $N_1$ is calculated by taking the ratio of the SM 
entropies after and before its decay
\begin{equation}
d \equiv \frac{s(\tilde T)}{s(T_d)} = 
\left( \frac{45 \zeta(3) \kappa_1^3}{\pi^4 \gs} \frac{M_1}{T_d} + 1 \right)^{3/4}.  
\label{eq:dilution}
\end{equation}

We will define the \emph{efficiency} $\eta$ to quantify the gain from hotness and the loss
from dilution as follows
\begin{equation}
\eta \equiv \frac{\kappa_1^3}{d} = 
\left( \frac{45 \zeta(3)}{\pi^4 \gs} \frac{M_1}{\kappa_1 T_d} 
+ \frac{1}{\kappa_1^4} \right)^{-3/4}\,.  
\label{eq:efficiency}
\end{equation}
We can derive the bounds on $\eta$ as follows. The upper bound is 
obtained by taking the largest $T_d$ from eq.~\eqref{eq:outofeq}:
\begin{equation}
\max (\eta) = \left( \frac{45 \zeta(3)}{\pi^4 \gs}
+ \frac{1}{\kappa_1^4} \right)^{-3/4}.
\label{eq:eta_max}
\end{equation}
For $\gs = 106.75$ and $\kappa_1 \gg 1$, we have $\max (\eta) \to 51.6$.
Its lower bound is obtained by taking the absolute minimum $T_d$ 
from eq.~\eqref{eq:reactivated_bound} and the largest $M_1$ from 
eq.~\eqref{eq:N1_relativistic}:
\begin{equation}
\min (\eta) = \left( \frac{45 \zeta(3)}{\pi^4 \gs} \frac{M_X}{z'_{\rm FO} T_{\rm EWsp}} 
+ \frac{1}{\kappa_1^4} \right)^{-3/4}.
\label{eq:eta_min}
\end{equation}

Next we will consider the following possible scenarios depending on 
the mass spectra of $N_i$. Denoting $T'_{\rm RH}$ as the reheating temperature in the dark sector after inflation, 
we have:
\begin{enumerate}

\item[$(i)$] {\bf $\boldsymbol{M_{2,\,3} \gg T'_{{\rm RH}} > M_{X}>M_{1}}$.}\\
In this case, both $N_{2}$ and $N_{3}$ are not generated in the thermal bath.
Notice that one can have arbitrary $\kappa_1 = \kappa_{\rm FO}$, taken 
as an initial condition; the upper bound on $m_l$ will be given by 
eq.~\eqref{eq:ml_bound}.

\item[$(ii)$] {\bf $\boldsymbol{M_{3} \gg T'_{{\rm RH}} > M_{X}>M_{2}>M_{1}}$.}\\
In this case, $N_{3}$ is not generated in the thermal bath. 
After the DM freeze-out, $N_2$ and $N_1$ have a temperature
$T'=\kappa_{\rm FO}\, T$, with $Y_{N_2}$ and $Y_{N_1}$ conserved. 
As seen in the previous section, $N_2$ decays when it is
still relativistic, heating up the SM sector. Again, assuming an instantaneous decay of $N_2$,
the energy conservation before and after its decay leads to
\begin{eqnarray}
\rho_{N_2}+\frac{\pi^{2}}{30}\gs\,T^{4} 
& = & \frac{\pi^{2}}{30}\gs \tilde{T}^{4}\,,\\
\frac{7\pi^{2}}{120}T'^{4}+\frac{\pi^{2}}{30}\gs\frac{T'^{4}}{\kappa_{\rm FO}^{4}} 
& = & \frac{\pi^{2}}{30}\gs\tilde{T}^{4}\,,
\end{eqnarray}
where $\tilde{T}$ is the SM temperature after the entropic injection from the decay of $N_2$.
Solving for $\tilde{T}$, we have\footnote{Notice that we would obtain the same result if we 
assume $N_2$ equilibrates with the SM sector first and then decays.}
\begin{equation}
\tilde{T} = \frac{T'}{\kappa_{\rm FO}}
\left(\frac{\frac{7}{4}\kappa_{\rm FO}^{4}+\gs}{\gs}\right)^{1/4}.
\end{equation}
Hence the ratio of the temperature of $N_1$ to the SM sector
will be
\begin{equation}
\kappa_1 = \frac{T'}{\tilde T}
=\kappa_{\rm FO}\left(\frac{\gs}{\frac{7}{4}\kappa_{\rm FO}^{4}+\gs}\right)^{1/4},
\label{eq:kappap}
\end{equation}
which is always smaller than $\kappa_{\rm FO}$.
In the limit of $\kappa_{\rm FO} \gg 1$, we have $\kappa_1\to\left(\frac47\, \gs\right)^{1/4}$
which gives $\kappa_1 \to 2.79$ with $\gs=106.75$. 
In this limit, from eq.~\eqref{eq:ml_bound}, we obtain $m_l < 1.6 \times 10^{-4}$ eV 
while from eq.~\eqref{eq:eta_max}, we have $\max(\eta) = 17.7$.

\item[$(iii)$] {\bf $\boldsymbol{T'_{{\rm RH}} > M_{X} > M_{3}>M_{2}>M_{1}}$.}\\
Similarly, using the approximation of instantaneous decays of $N_2$ and $N_3$, 
the ratio of the temperature of $N_1$ to the SM sector is 
\begin{equation}
\kappa_1 = \kappa_{\rm FO}\left(\frac{\gs}{\frac{7}{2}{\kappa_{\rm FO}}^{4}+\gs}\right)^{1/4}.
\label{eq:kappapp}
\end{equation}
In the limit of $\kappa_{\rm FO}\gg 1$, 
we have $\kappa_1 \to\left(\frac 27\, \gs\right)^{1/4}$ 
which takes the value $2.35$ with $\gs = 106.75$. 
In this limit, from eq.~\eqref{eq:ml_bound}, we obtain $m_l < 2.1 \times 10^{-4}$ eV
while from eq.~\eqref{eq:eta_max}, we have $\max(\eta) = 11.6$.
\end{enumerate}

\begin{figure}
\begin{centering}
\includegraphics[width=0.328\textwidth]{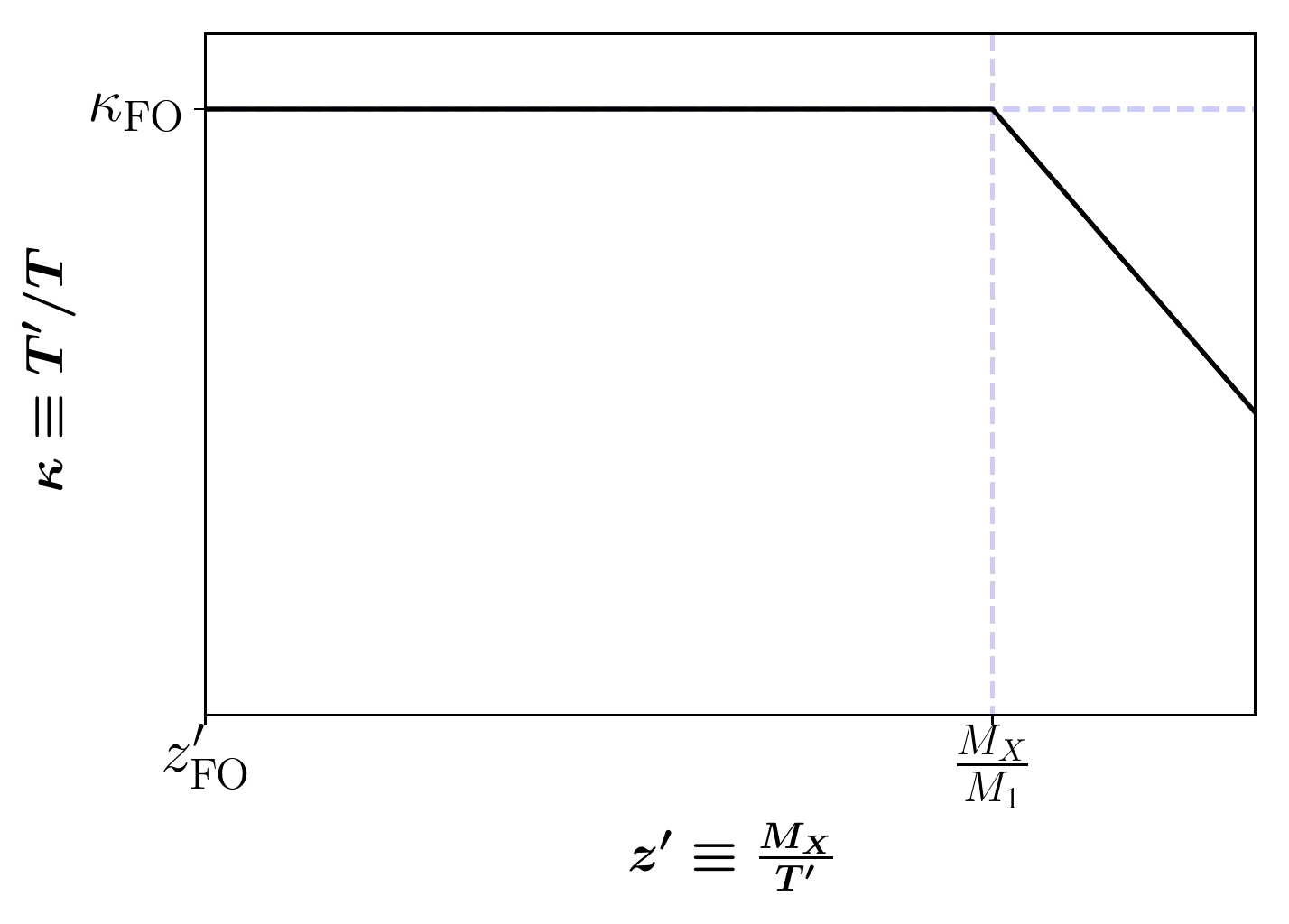}
\includegraphics[width=0.328\textwidth]{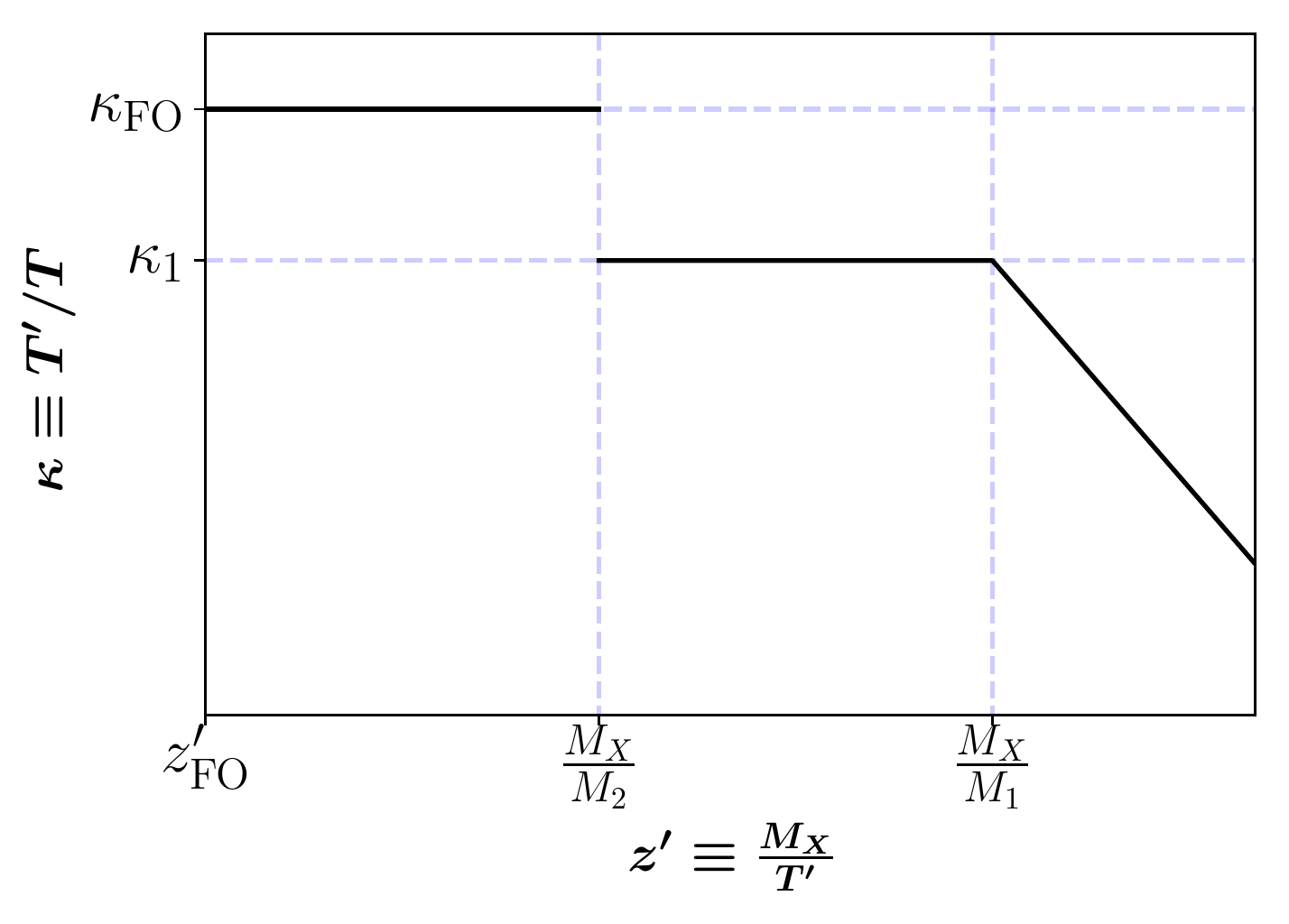}
\includegraphics[width=0.328\textwidth]{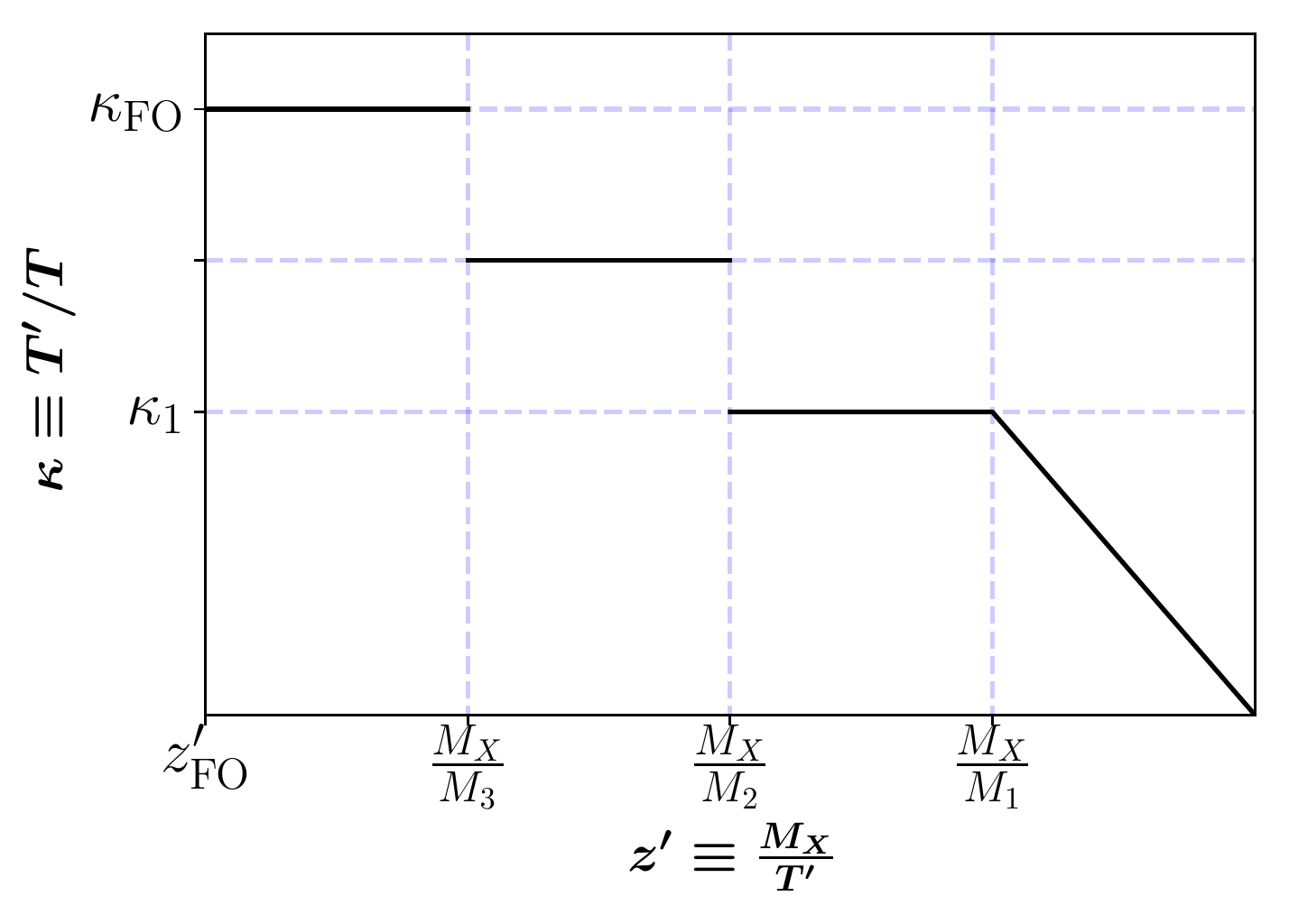}
\par\end{centering}
\caption{
Here we illustrate the ratio of the temperature of $N_1$ to the SM sector, $\kappa_1$
for three possible scenarios. In scenario $(i)$ with $N_2$ and $N_3$ much heavier than $T_{\rm}$, 
they are not produced. In this case, $\kappa_1 = \kappa_{\rm FO}$ 
will remain constant until when $N_1$ becomes nonrelativistic, when it decays. 
In scenario $(ii)$, $\kappa$ decreases due to decay of $N_2$ at $T' = M_2$ which only heats 
up the SM sector. For scenario $(iii)$, $\kappa$ decreases twice due to decays of both $N_3$ and $N_2$ 
at $T' = M_3$ and $T' = M_2$, respectively.
}
\label{fig:toy}
\end{figure}

Figure~\ref{fig:toy} illustrates the temperature evolution of $N_1$ of the three possible scenarios discussed above.


\subsection{Scatterings, Oscillations, and Coherence of the RHNs}
\label{sec:scatt_osc}

In the above, we have ignored the possibility that the RHNs 
produced from DM annihilations are in general not mass
eigenstates but some arbitrary quantum states $\hat{N}_{a}$ $(a,b=1,\,2,\,3)$
which are superpositions of the mass eigenstates $N_{i}$. Hence it
is possible for oscillations $\hat{N}_{a} \leftrightarrow \hat{N}_{b}$ to occur efficiently.
Notice that for scenario $(i)$, only $N_{1}$ will be produced from
$XX\to N_{1}N_{1}$ and hence hot leptogenesis can always proceed.
However, this is not guaranteed in scenarios $(ii)$ nor $(iii)$. 
For scenario $(ii)$, the quantum states will be the superposition of 
only $N_{1}$ and $N_{2}$ while for scenario $(iii)$, 
of all three $N_{1}$, $N_{2}$ and $N_{3}$.
In such cases, one has to take into account scatterings of $\hat N_a$ with the SM particles,
the oscillations of $\hat N_a$ and possible decoherence of $\hat N_a$.

\subsubsection{Scatterings}
First of all, we have to make sure that the scatterings of $\hat N_a$ 
with the SM particles are only in equilibrium after the DM freeze-out.
Otherwise, the temperatures of both $\hat{N}_{a}$ and the SM sectors will equalize,
making the scenarios $(ii)$ and $(iii)$ not viable. At $T'>M_{i}$,
the scatterings of $\hat{N}_{a}$ with the SM particles can be estimated
similar to refs.~\cite{Cline:1993bd,Garbrecht:2014kda}, as follows
\begin{eqnarray}
\Gamma_{a}^{{\rm scatt}} & \approx & 
5\times10^{-3}\left(\hat{\lambda}^{\dagger}\hat{\lambda}\right)_{aa}T'\nonumber \\
& \approx & 5\times10^{-3}\sum_{i,j}W_{ia}^{*}W_{ja}
\frac{\left(R^{\dagger}\hat{m}R\right)_{ij}\sqrt{M_{i}M_{j}}}{v^{2}}\,T'\,,
\label{eq:Gamma_scatt_gen}
\end{eqnarray}
where $\hat{\lambda}\equiv\lambda\,W$ with $W$ being a unitary matrix associated
with the basis $\hat{N}_{a}=\sum_{i}W_{ia}^{*}\,N_{i}$ produced from
DM annihilations; in the last line, we have made use of eq.~\eqref{eq:Casas_Ibarra_Param}.
We further approximate the scattering rate by setting 
$M_N \equiv \sqrt{M_i M_j}$ to be a common scale and obtain
\begin{equation}
\Gamma_{a}^{{\rm scatt}} \approx 5\times10^{-3}
\frac{\left(W^\dagger R^{\dagger}\hat{m}R W\right)_{aa} M_N}{v^{2}}\,T'\,.
\label{eq:Gamma_scatt}
\end{equation}
Comparing eq.~\eqref{eq:Gamma_scatt} to the Hubble rate, eq.~\eqref{eq:Hubble}, 
we have
\begin{eqnarray}
K_{a}^{{\rm scatt}}\equiv \frac{\Gamma_{a}^{{\rm scatt}}}{H} 
& \approx & 1.2\times10^{2} 
\frac{ (W^\dagger R^\dagger \hat{m} R W)_{aa} }{0.1\,{\rm eV}}\, \frac{M_N}{T'}
\frac{1}{\sqrt{\gsp+\frac{\gs}{\kappa^{4}}}}\,.
\label{eq:K_scatt}
\end{eqnarray}

Depending on scenarios $(ii)$ or $(iii$) and also the mass ordering, 
we can derive the bound on $(W^\dagger R^\dagger \hat{m} R W)_{aa}$. 
For scenario $(ii)$, we have
\begin{eqnarray}
(W^\dagger R^\dagger \hat{m} R W)_{aa}  & \geq & \begin{cases}
\frac{1}{2} \sqrt{\Delta m_{\rm sol}^2} & \;\;\;\mbox{for NO},\\
\frac{1}{2} \sqrt{\Delta m_{\rm atm}^2} & \;\;\;\mbox{for IO}.
\end{cases}
\end{eqnarray}
For scenario $(iii)$, we have
\begin{eqnarray}
(W^\dagger R^\dagger \hat{m} R W)_{aa}  & \geq & \begin{cases}
\frac{1}{3} \sqrt{\Delta m_{\rm atm}^2} & \;\;\;\mbox{for NO},\\
\frac{2}{3} \sqrt{\Delta m_{\rm atm}^2} & \;\;\;\mbox{for IO}.
\end{cases}
\end{eqnarray}

From eq.~\eqref{eq:K_scatt}, the scatterings will get into equilibrium when
\begin{equation}
T' \lesssim T'_\text{scatt}\equiv 61 \frac{ (W^\dagger R^\dagger \hat{m} R W)_{aa} }{\sqrt{\Delta m_{\rm atm}^2}}
\frac{M_N}{\sqrt{\gsp+\frac{\gs}{\kappa^{4}}}}\,.
\label{eq:scatt_eq}
\end{equation}
To make sure that the scatterings only get into equilibrium after 
the DM freeze-out, we require $ T'_\text{scatt} < T'_{\rm FO}$. 
This gives
\begin{equation}
z'_{\rm FO} \equiv \frac{M_X}{T'_{\rm FO}}  \lesssim  
0.016 \frac{\sqrt{\Delta m_{\rm atm}^2}}{ (W^\dagger R^\dagger \hat{m} R W)_{aa} }
\sqrt{\gsp+\frac{\gs}{\kappa_{\rm FO}^{4}}} \frac{1}{r}\,,
\label{eq:scatt_constraint}
\end{equation}
where we have set $\kappa = \kappa_{\rm FO}$ and defined
\begin{equation}
r \equiv \frac{M_N}{M_X}.
\label{eq:ratio_MN_MX}
\end{equation}

We can rewrite the constraint above as 
\begin{equation}
M_1 < r M_X  \lesssim  
0.016 \frac{\sqrt{\Delta m_{\rm atm}^2}}{ (W^\dagger R^\dagger \hat{m} R W)_{aa} }
\sqrt{\gsp+\frac{\gs}{\kappa_{\rm FO}^{4}}} \frac{M_X}{z'_{\rm FO}},
\label{eq:scatt_constraint_hierarchy}
\end{equation}
in which the strongest constraint is obtained by taking $\kappa_{\rm FO} \gg 1$. 

In our study, we take $\Gamma^{{\rm ann}}$ to be 
\begin{equation}
\Gamma^{{\rm ann}}  \equiv  n_{X}\left\langle \sigma v_{X}\right\rangle 
=g_{X}\left(\frac{M_{X}^{2}}{2\pi\, z'}\right)^{3/2}e^{-z'}\frac{4\pi\,\xi}{v_{X}M_{X}^{2}}
=\sqrt{\frac{2}{\pi}}\frac{\xi\, g_{X}\,e^{-z'}}{v_{X}\,{z'}^{3/2}}M_{X}\,,
\label{eq:ann_rate}
\end{equation}
where $g_{X}$ denotes the degrees of freedom of $X$ and $z'\equiv\frac{M_{X}}{T'}$.
In the above, we have assumed $M_{X}\gg M_{i}$ and parametrize the
annihilation cross section as
\begin{equation}\label{eq:maxsv}
\left\langle \sigma v_{X}\right\rangle  = \frac{4\pi\,\xi}{v_{X}M_{X}^{2}}\,,
\end{equation}
where $\xi\leq1$, with $\xi=1$ corresponding to the maximum cross section bounded
by unitarity~\cite{Griest:1989wd,Hui:2001wy}.

Another relevant scattering process that we should take into account
is $\hat{N}_{a}\hat{N}_{b}\leftrightarrow\hat{N}_{c}\hat{N}_{d}$
mediated by $X$ in the loop. But due to the divergence in the loop,
this result does not make sense and we need an underlying ultraviolet
theory. As an example, we will consider a simple ultraviolet complete
model to illustrate that it is possible to suppress $\hat{N}_{a}\hat{N}_{b}\leftrightarrow\hat{N}_{c}\hat{N}_{d}$
while having a large $XX\to\hat{N}_{a}\hat{N}_{b}$. We introduce
a scalar particle $\phi$ with the following coupling
\begin{equation}
-{\cal L}  = y_{ab}\hat{N}_{a}\hat{N}_{b}\phi+y_{X}XX\phi+{\rm H.c.}\,.
\end{equation}
In this model, we can suppress $\hat{N}_{a}\hat{N}_{b}\leftrightarrow\hat{N}_{c}\hat{N}_{d}$
by choosing $y_{X}\lesssim10^{-3}$ while taking $M_{\phi}\sim2M_{X}$
to have large annihilation rate $XX\to\hat{N}_{a}\hat{N}_{b}$. 

\subsubsection{Oscillations}\label{sec:osc}
Next let us estimate the oscillation rate $\Gamma^{{\rm osc}}$
as follows
\begin{equation}
\Gamma^{{\rm osc}} = \frac{M_{j}^{2}-M_{1}^{2}}{4\pi\,E}
\equiv \frac{\delta_j\, M_j^2}{12.6\pi\, T'},
\end{equation}
where we have taken $E=3.15\, T'$ the energy for relativistic
fermion \cite{Kolb:1990vq} and defined the mass splitting parameter as
\begin{equation}
\delta_{j} \equiv 1 - \frac{M_1^2}{M_j^2}\,,
\label{eq:delta_j}
\end{equation}
with $j=2$ or $3$. 
Comparing to the Hubble rate eq.~\eqref{eq:Hubble}, we have
\begin{equation}
K^{{\rm osc}}\equiv\frac{\Gamma^{{\rm osc}}}{H} 
 =  \frac{\left(M_{j}^{2}-M_{1}^{2}\right)M_{{\rm Pl}}}
{\sqrt{\frac{4\pi^{3}}{45}\left(\gsp+\frac{\gs}{\kappa^{4}}\right)}\,T'^{3}}
=\frac{\delta_{j}\,M_{{\rm Pl}}\,{z}_{j}^{'3}}
{\sqrt{\frac{4\pi^{3}}{45}\left(\gsp+\frac{\gs}{\kappa^{4}}\right)}\,M_{j}}\,,
\end{equation}
where we have defined $z'_{i}\equiv\frac{M_{i}}{T'}$. 
Taking $\gsp=6$ and $\kappa\gg1$, we have
\begin{equation}
K^{{\rm osc}} =3\times10^{9}\left(\frac{z'_{j}}{10^{-1}}\right)^{3}
\left(\frac{\delta_{j}}{10^{-3}}\right)\left(\frac{1\,{\rm TeV}}{M_{j}}\right),
\label{eq:K_osc}
\end{equation}
which implies that the oscillations are very fast at early time even
with a very small mass splitting. 

\begin{figure}
\begin{centering}
\includegraphics[width=0.47\textwidth]{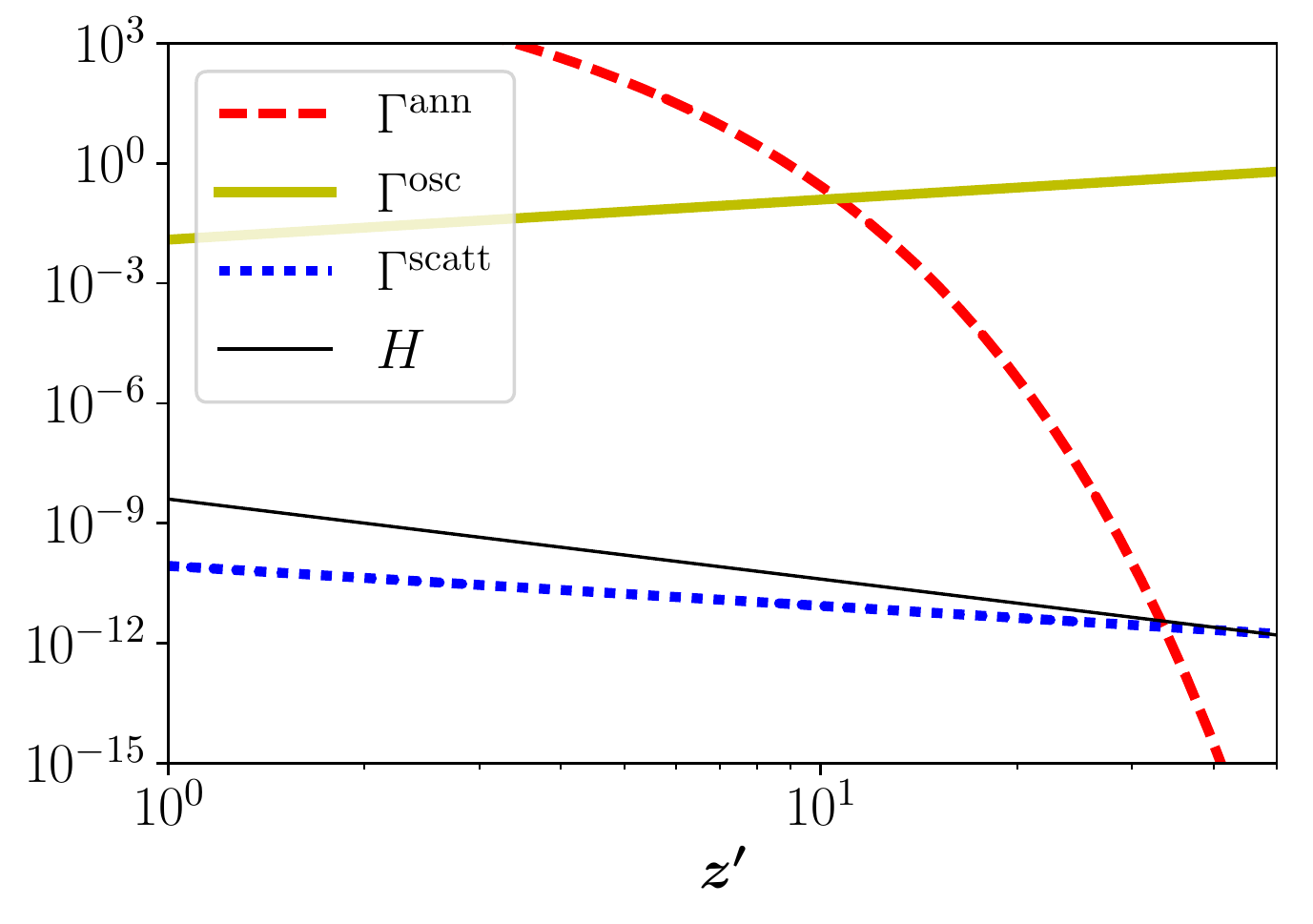}
\par\end{centering}
\caption{Here we plot $\Gamma_a^{\rm ann}$ (red dashed), $\Gamma^{\rm scatt}$ (blue dotted)
and $\Gamma^{{\rm osc}}$ (green solid) as functions of $z'$ with
the following parameters: $v_{X}=0.425$, $\xi=1$,
$(W^\dagger R^\dagger \hat{m} R W)_{aa} = \sqrt{\Delta m_{\rm atm}^2}/3$, 
$M_j = M_N$ with $r=320$ and $M_{X}=10^5$ GeV and $g_{X}=1$
with $\delta_{j}=0.5 $. We also plot the Hubble rate (black solid) with $g_\star' = 6$ 
and $\kappa = 2.35$.}
\label{fig:rate}
\end{figure}

The constraint~\eqref{eq:scatt_constraint} will be applied in our numerical analysis. 
For illustration, in figure~\ref{fig:rate}, we plot 
$\Gamma_a^{\rm ann}$ (red dashed), $\Gamma^{\rm scatt}$ (blue dotted)
and $\Gamma^{{\rm osc}}$ (green solid) as functions of $z'$ with 
the following parameters: $v_{X}=0.425$, $\xi=1$, 
$(W^\dagger R^\dagger \hat{m} R W)_{aa} = \sqrt{\Delta m_{\rm atm}^2}/3$, 
$M_j = M_N$ with $r=320$ and $M_{X}=10^5$ GeV and $g_{X}=1$
with $\delta_{j}=0.5 $ in comparison to the Hubble rate (black solid) with $g_\star' = 6$ 
and $\kappa = 2.35$.

\subsubsection{Coherence}\label{sec:coh}
In estimating the oscillation rate above, we have ignored possible 
decoherence of $\hat N_a$ generated from the annihilations $X\,X \to \hat N_a\, \hat N_b$.
In fact, $\hat{N}_{a}\leftrightarrow\hat{N}_{b}$ oscillations can 
only happen if the produced $\hat{N}_{a}$ remain coherent superpositions of $N_{i}$. 
In particular, once the $\hat{N}_{a}$ decoheres, they will be projected to 
mass eigenstates $N_{i}$ and oscillations cease to occur. 
If decoherence happens before $\hat{N}_{a}$ scatterings
with the SM particles can take place, $N_{1}$, being out of equilibrium,
will remain hotter than the SM and $N_{2}$ ($N_3$).
Next we turn to the study
of coherence of the quantum state $\hat{N}_{a}$. There are two possible
mechanisms of decoherence akin to those which happen in neutrino oscillations~\cite{Giunti:1997wq,Fong:2016yyh}
that will ensure $N_{1}$ remains hotter than the SM sector:
\begin{enumerate}
\item[$(a)$] If the $\hat{N}_{a}$ produced from DM annihilations are
in mass eigenstates $N_{i}$, they do not oscillate. For this to happen,
the uncertainty in the energy of $\hat{N}_{a}$ has to be smaller
than the energy differences of $N_i$ due to their mass differences, in
which case, the different eigenstates can be distinguished:\footnote{In particular, we are interested in the case where mass eigenstate $N_1$ is projected out due to the decoherence.}
\begin{equation}
\delta E  <  \left|E_{j}-E_{1}\right|\approx\frac{M_{j}^{2}-M_{1}^{2}}{2E}
= 2\pi\, \Gamma^{{\rm osc}}.
\label{eq:phy_condition_a}
\end{equation}
We can estimate $\delta E$ from the size of the wavepacket of $N_{i}$.
From the uncertainty principle $\delta p_{a}\,\delta x\approx\delta E_{a}\,\delta x\approx1$,
we obtain $\delta E\approx\delta x^{-1}$. Next we estimate
$\delta x$ to be the mean free path of $X$ before it annihilates
with another $X$ to produce a pair of $\hat{N}_{a}$ as follows
\begin{equation}
\delta x = v_{X}\,t^{{\rm ann}}=\frac{v_{X}}{\Gamma^{{\rm ann}}}\,,
\label{eq:delta_x}
\end{equation}
with $v_{X}$ the relative velocity of $X$ and $\Gamma^{{\rm ann}}$ 
is given by eq.~\eqref{eq:ann_rate}.

Using eq.~\eqref{eq:delta_x}, we have 
\begin{equation}
\delta E \approx \delta x^{-1}
=\sqrt{\frac{2}{\pi}}\frac{\xi\, g_{X}\,e^{-z'}}{v_{X}^{2}\,{z'}^{3/2}}M_{X}\,.
\label{eq:delta_E}
\end{equation}
Finally, from eq.~\eqref{eq:phy_condition_a}, we obtain the condition
\begin{equation}
d^{\rm (a)} \equiv \frac{\delta E}{2\pi}
=\frac{\xi\, g_{X}\,M_{X}}{\sqrt{2}\pi^{3/2}\,v_{X}^{2}}\,
{z'}^{-3/2}\,e^{-z'}  <  \Gamma^{{\rm osc}}.
\label{eq:decoh_a}
\end{equation}
Notice that eq.~\eqref{eq:decoh_a} might be fulfilled for low temperature.

\item[$(b)$] If the wavepackets of $N_{i}$ separate, $\hat{N}_{a}$ are projected
into respective mass eigenstates $N_{i}$ and oscillations cease to
happen. For this to arise, the uncertainty in the energy of $\hat{N}_{a}$
has to be sufficiently large such that the wavepackets of $N_{i}$ are narrow
enough and separate before they scatter with the SM particles. The
distance traveled while $\hat{N}_{a}$ remains coherent can be estimated
to be~\cite{Giunti:1997wq}
\begin{equation}
x^{{\rm coh}}  \equiv  \frac{E}{\delta E}\, \frac{v_{N}}{\Gamma^{{\rm osc}}}\,,
\end{equation}
where $v_{N}$ is the velocity of $\hat{N}_{a}$. The condition such
that the wavepackets separate before they scatter translates into
\begin{equation}
x^{{\rm coh}} < v_{N}\,t^{{\rm scatt}}=\frac{v_{N}}{\Gamma_{a}^{{\rm scatt}}}\,.
\end{equation}
Putting all these together, finally we have
\begin{equation}
d^{\rm (b)}\equiv\frac{E}{\delta E}\,\Gamma_{a}^{{\rm scatt}} 
 <  \Gamma^{{\rm osc}}\,.
\label{eq:decoh_b}
\end{equation}
Using $E=3.15\,T'$, eqs.~\eqref{eq:Gamma_scatt} and~\eqref{eq:delta_E}, one gets
\begin{equation}
d^{\rm (b)} \approx 2.0\times 10^{-2}\, 
(W^\dagger R^\dagger \hat{m} R W)_{aa}\, 
\frac{v_{X}^{2}\, e^{z'}}{\xi\, g_{X}\, {z'}^{1/2}} \frac{M_N\, M_{X}}{v^2}\,.
\end{equation}
Notice that eq.~\eqref{eq:decoh_b} might be fulfilled for high temperature.
\end{enumerate}

\begin{figure}
\begin{centering}
\includegraphics[width=0.47\textwidth]{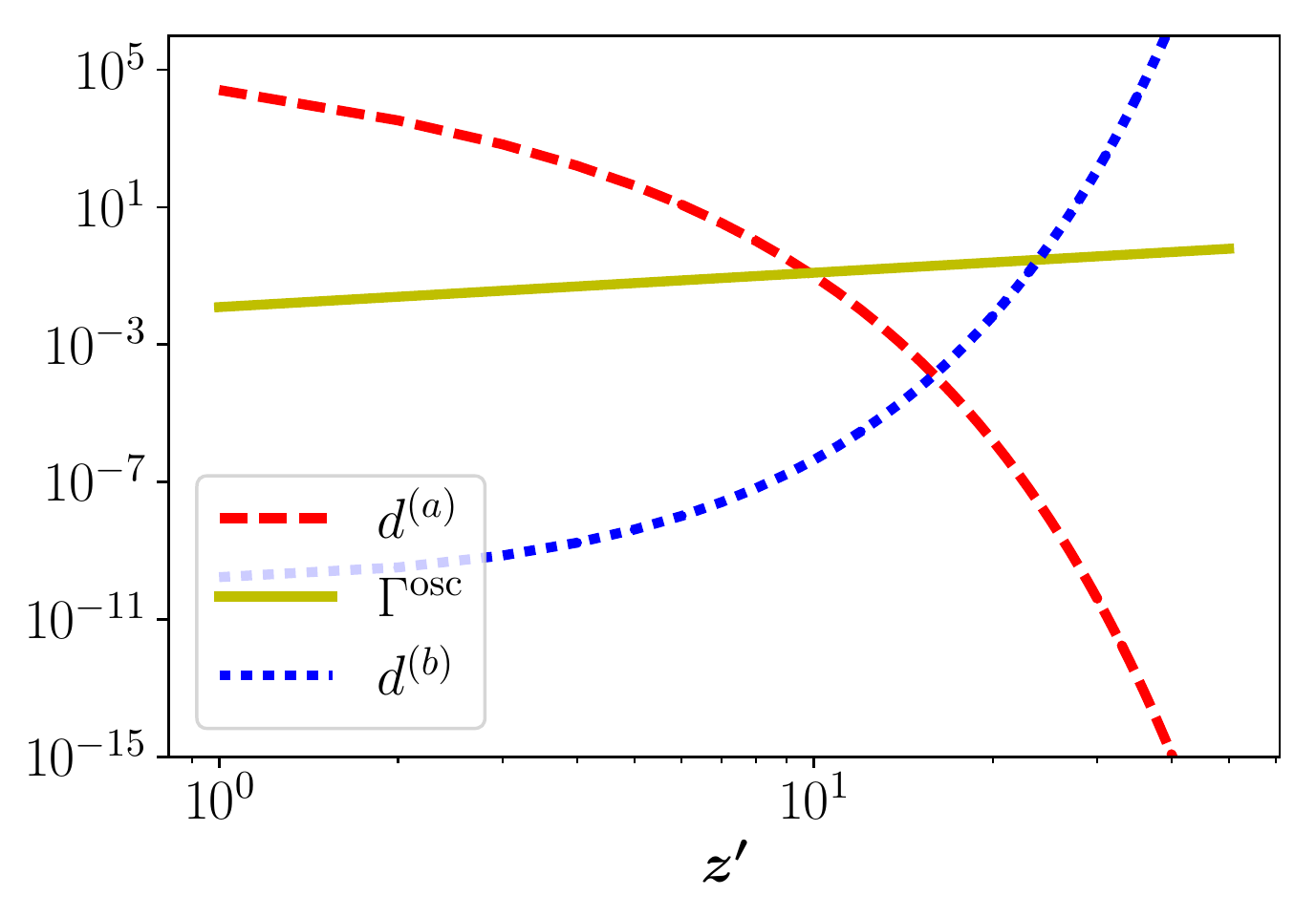}
\includegraphics[width=0.47\textwidth]{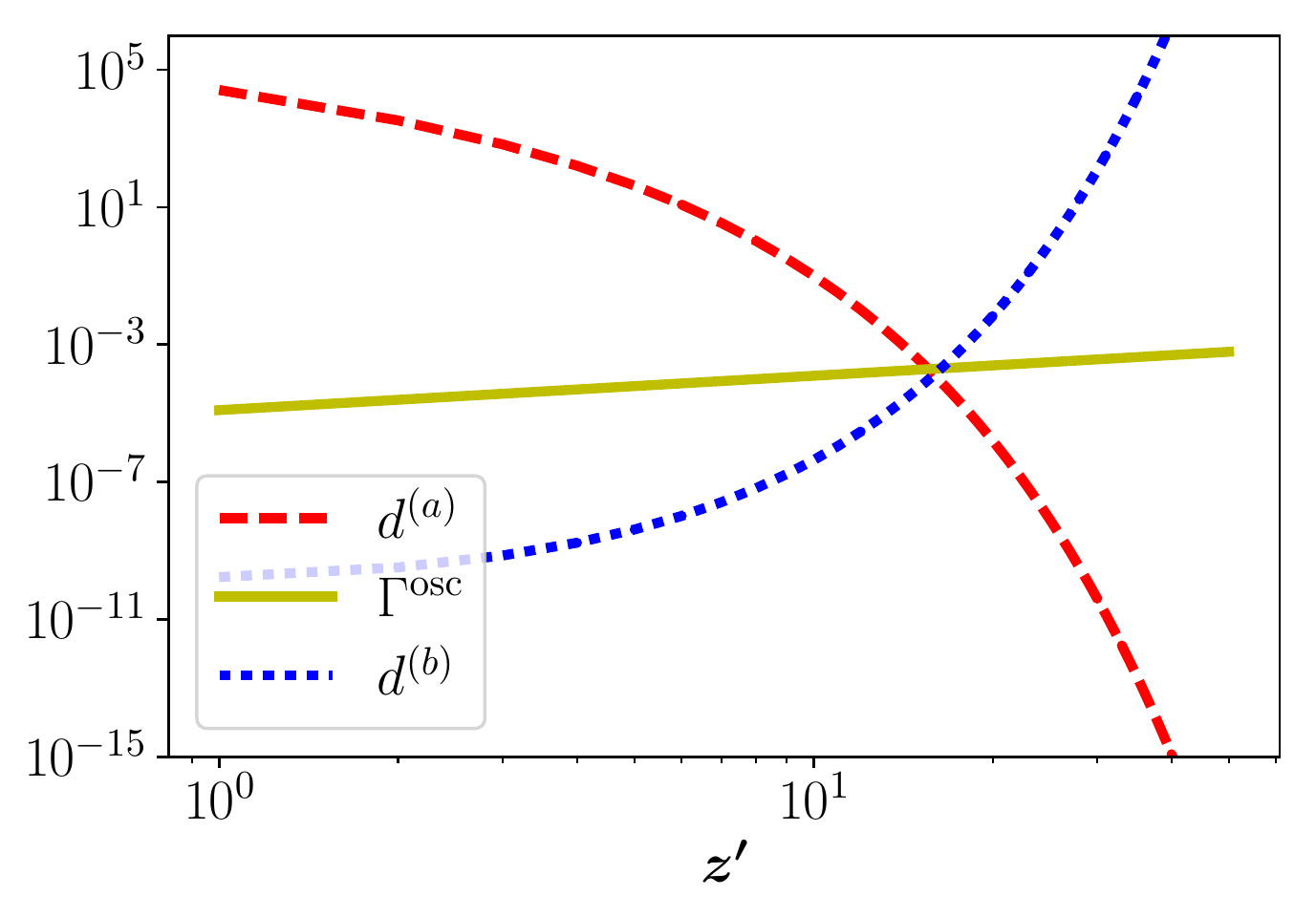}
\par\end{centering}
\caption{Here we plot $d^{\rm (a)}$ (magenta dotted), $d^{\rm (b)}$ (purple dashed)
and $\Gamma^{{\rm osc}}$ (green solid) as functions of $z'$ with
the following parameters: $v_{X}=0.425$, $\xi=1$,
$(W^\dagger R^\dagger \hat{m} R W)_{aa} = \sqrt{\Delta m_{\rm atm}^2}/3$, 
$M_N = M_{j}$ with $r=320$ and $M_{X}=10^5$ GeV and $g_{X}=1$
with two choices of $\delta_{j}=\left\{0.5,5\times 10^{-4}\right\} $
(left, right). In all the cases,
the decoherence conditions $d^{(a)} < \Gamma^{\rm osc}$ eq.~\eqref{eq:decoh_a} 
and/or $d^{(b)} < \Gamma^{\rm osc}$ eq.~\eqref{eq:decoh_b}
are fulfilled until the freeze-out of $XX\to NN$ at typically
about $z_{X}\gtrsim 30$. }
\label{fig:decoherence}
\end{figure}

For scenarios $(ii)$ and $(iii)$, the condition \eqref{eq:decoh_a} 
\emph{or} \eqref{eq:decoh_b} needs to be fulfilled \emph{at all time}.
They can be violated for very degenerate $N_j$, i.e.
$\delta_j \ll 1$, and a lower bound on $\delta_j$ is obtained when 
$d^{\rm (a)} = d^{\rm (b)} = \Gamma^{\rm osc}$.
For illustration, in figure~\ref{fig:decoherence}, 
we plot the three quantities $d^{\rm (a)}$ , $d^{\rm (b)}$
and $\Gamma^{{\rm osc}}$ as functions of $z'$ by choosing the following
parameters: $v_{X}=0.425$, $\xi=1$, 
$(W^\dagger R^\dagger \hat{m} R W)_{aa} = \sqrt{\Delta m_{\rm atm}^2}/3$, $M_N = M_{j}$
with $r=320$ and $M_{X}=10^5$ GeV and $g_{X}=1$ with two choices
of $\delta_{j}=\left\{0.5,5\times 10^{-4}\right\} $.

In this section, we have shown that naively $\hat{N}_{a}\leftrightarrow\hat{N}_{b}$
oscillations and $\hat{N}_{a}$ scatterings with the SM particles
are in thermal equilibrium and this will equilibrate the temperature
of both sectors making hot leptogenesis for scenarios $(ii)$ and $(iii)$ not viable. 
Nevertheless, as long as we make sure that the scatterings only get into 
equilibrium after the freeze-out of $XX \to NN$ (condition \eqref{eq:scatt_constraint_hierarchy})
and that decoherence is efficient (conditions \eqref{eq:decoh_a} and \eqref{eq:decoh_b}),
$N_1$ responsible for leptogenesis can remain hotter than $N_2$ and $N_3$ and
scenarios $(ii)$ or $(iii)$ will remain viable.

Finally, we would like to point out that the decays of $N_i$ 
can in principle also lead to decoherence. Nevertheless, due to 
time dilation at $T' \gg M_i$, the effective decay rate is given by\footnote{The thermal 
average time dilation factor is given by 
$\left\langle\frac{M_i}{E}\right\rangle = \frac{{\cal  K}_1(z_i')}{{\cal K}_2(z_i')}$ 
where ${\cal K}_{n}$ is modified Bessel function of second kind of order-$n$ and $z_i' \equiv M_i/T'$.}
\begin{equation}
\Gamma_i^{\rm eff} = \left\langle\frac{M_i}{E}\right\rangle \Gamma_i 
\simeq \frac{z_i'}{2} \,\Gamma_i\,,
\end{equation}
where in the last step, we have expanded in $z_i' \ll 1$.
Since DM freezes out when $N_i$ is still relativistic, 
the decays are slow compared to the Hubble rate and this effect is negligible.  


\subsection{Realizations of Leptogenesis}
\label{sec:realizations_leptogenesis}

In this section, we will discuss the realizations of 
leptogenesis. We will start by considering a hierarchical mass spectrum $N_i$.
In this case, leptogenesis proceeds through the decay of the lightest RHN, $N_{1}$.
For its decay, which is \emph{out-of-equilibrium}, the
baryon asymmetry $Y_{B}$ generated from an initial abundance $Y_{N_{1}}^0$ 
is given by\footnote{This formula is a very good description for 
$K_1 \ll 1$ while for $K \sim 1$, there will be suppression of order one factor from washout.}
\begin{equation}
Y_{B} = -\frac{28}{79}\,\epsilon \,Y_{N_{1}}^0\,,
\label{eq:baryon_asym}
\end{equation}
where $\epsilon$ is the CP violation parameter defined as
\begin{equation}
\epsilon \equiv 
\frac{\sum_\alpha \left[\gamma\left(N_{1}\to\ell_\alpha\phi\right)
-\gamma\left(N_{1}\to\overline{\ell}_\alpha\phi^{*}\right)\right]}
{\sum_\beta \left[\gamma\left(N_{1}\to\ell_\beta\phi\right)
+\gamma\left(N_{1}\to\overline{\ell}_\beta\phi^{*}\right)\right]}\,,
\label{eq:CP_param}
\end{equation}
where $\gamma\left(a\to b\right)$ is the thermally averaged reaction density.
In the above we have summed over lepton flavors because flavor 
effects~\cite{Barbieri:1999ma,Abada:2006fw,Nardi:2006fx,Abada:2006ea} 
are negligible the weak washout regime ($K_1 < 1$), for an initial abundance of $N_1$ 
which is dominantly generated from new interactions other than the neutrino Yukawa 
(as it is the case in this scenario, where the RHNs are populated through interactions within the dark sector). 
This result holds also at $T \lesssim 10^9$ GeV when lepton flavors can be distinguished 
in the thermal bath. 
The reason is the following. Without DM annihilations, $N_1$ is generated solely 
through inverse decays $\phi\, \ell_\alpha \to N_1$ and the $B/3-L_\alpha$ asymmetry ($L_\alpha$ being 
the lepton flavor number) generated during 
this period depends crucially on the flavor-dependent washout. The final $B-L$ 
asymmetry will be the sum of the flavor-dependent $B/3-L_\alpha$ asymmetries, 
generated during $N_1$ population, and the $B-L$ asymmetry generated during the decay of $N_1$,
when the flavor-dependent washout is negligible. 
On the other hand, if $N_1$ is populated mainly from new interaction 
like $X\,X\to N_i\,N_i$, the $B-L$ 
asymmetry will be generated dominantly from $N_1$ decays where the flavor-dependent 
washout is negligible. 
Since $N_1$ decays late at $T \ll M_1$, thermal effects~\cite{Covi:1997dr,Giudice:2003jh} 
also have a negligible impact.

At the leading order, the CP violation parameter of eq~\eqref{eq:CP_param} is given by~\cite{Covi:1996wh}
\begin{equation}
\epsilon = \frac{1}{8\pi\left(\lambda^{\dagger}\lambda\right)_{11}}
\sum_{j\neq1}{\rm Im}\left[\left(\lambda^{\dagger}\lambda\right)_{1j}^{2}\right]\,g\left(\frac{M_{_{j}}^{2}}{M_{1}^{2}}\right)\,,
\end{equation}
with
\begin{equation}
g\left(x\right) = \sqrt{x}\left[\frac{1}{1-x}+1-\left(1+x\right)\ln\left(\frac{1+x}{x}\right)\right]\,.
\end{equation}

Assuming hierarchical RHNs with $M_{1}\ll M_{2}<M_{3}$, the CP
parameter can be rewritten as
\begin{equation}
\epsilon = -\frac{3M_{1}}{16\pi v^{2}}\frac{{\rm Im}\left[\left(\lambda^{\dagger}m_{\nu}\lambda^{*}\right)_{11}\right]}{\left(\lambda^{\dagger}\lambda\right)_{11}}\,,
\end{equation}
where we have used eq.~\eqref{eq:typeI_nu_mass}.
With Casas-Ibarra parametrization of eq.~\eqref{eq:Casas_Ibarra_Param},
we obtain
\begin{equation}
\epsilon = -\frac{3M_{1}}{16\pi\, v^{2}}
\frac{\sum_{k}m_{k}^{2}\,{\rm Im}\left(R_{k1}^{*2}\right)}{\sum_{i}m_{i}\left|R_{i1}\right|^{2}}\,.
\end{equation}
From the above, we can derive the Davidson-Ibarra bound~\cite{Davidson:2002qv}
\begin{equation}
\left|\epsilon\right| \leq \epsilon_{{\rm DI}}
=\frac{3M_{1}}{16\pi v^{2}}\left(m_{h}-m_{l}\right)
=\frac{3M_{1}}{16\pi v^{2}} \frac{\Delta m_{\rm atm}^2}{m_{h}+m_{l}}\,,
\label{eq:DI_bound}
\end{equation}
where $m_{h}$ ($m_{l}$) denotes the heaviest (lightest) light neutrino
mass. Substituting eq.~\eqref{eq:DI_bound} into eq.~\eqref{eq:baryon_asym}
and requiring $\left|Y_{B}\right|\geq Y_{B}^{{\rm obs}}$, we have
\begin{equation}
M_{1} \geq \frac{316}{21}\,\frac{v^{2}(m_h + m_l)}{\Delta m_{\rm atm}^2}\,
\frac{Y_{B}^{{\rm obs}}}{Y_{N_{1}}^0}\,.
\end{equation}
Taking $m_{h}+m_{l}>\sqrt{\Delta m_{{\rm atm}}^{2}}$ and 
$Y_{B}^{{\rm obs}}=8.7\times10^{-11}$~\cite{Ade:2015xua},
we have
\begin{equation}
M_{1} > \frac{7.9\times10^{5}}{Y_{N_{1}}^0}\,{\rm GeV}.
\label{eq:N1_bound}
\end{equation}

In this scenario, one has to take into account of dilution 
due to the late decay of $N_1$, eq.~\eqref{eq:dilution},
and setting $Y_{N_1}^0 = Y_{N_1}$ from eq.~\eqref{eq:N1_abundance}, 
the previous bound becomes
\begin{equation}
M_{1} \geq 7.9\times 10^{5}~{\rm GeV}\,\frac{4\pi^{4}\, \gs} {135\zeta(3)\, \eta}
\simeq 2.0\times 10^8~\text{GeV}\,\left(\frac{\gs}{106.75}\right)\,\frac{1}{\eta}\,,
\label{eq:N1_bound_hot}
\end{equation}
where $\eta$ is given by eq.~\eqref{eq:efficiency}. 
The absolute lower bound is obtained by considering the case $\kappa_1 \gg 1$, 
which gives $\max(\eta) \to 51.6$, and corresponds to $M_{1} \geq 3.9\times 10^{6}$ GeV.
We can compare this bound to the result of ref.~\cite{Giudice:2003jh} 
which obtained $M_1 > 1.7 \times 10^7$ GeV for the case where $N_1$ 
dominates the energy density of the Universe at early times.

As we will see in the next section, the absolute lower mass bound derived above for hierarchical 
RHNs \eqref{eq:N1_bound_hot} is in tension with having the correct DM relic abundance and 
relativistic $N_1$ during freeze-out which requires $M_1 \lesssim 4 \times 10^3\,{\rm GeV}$.
For scenarios $(ii)$ and $(iii)$, the bound can be relax once we allow 
for quasi-degenerate $N_i$ to resonantly enhance the CP violation \cite{Pilaftsis:1997jf}. 
From eq.~\eqref{eq:N1_bound_hot} and $M_1 \lesssim 4 \times 10^3\,{\rm GeV}$, 
the mass splitting (eq.~\eqref{eq:delta_j}) we need is estimated 
to be\footnote{In the case where the mass splitting $\delta_j M_j^2$ is 
of the order of decay width $\Gamma_j$, one will need to include the regulator obtained from 
the resummed heavy neutrino propagator as in ref.~\cite{Pilaftsis:1997jf}. 
In our case, however, we have $\delta_j M_j^2 \gg \Gamma_j$ and the effect of 
including the regulator is negligible and we can estimate the maximal CP parameter 
\eqref{eq:DI_bound} to be enhanced by $1/\delta_j$.}
\begin{equation}
\delta_{j} \lesssim 2\times 10^{-5}~{\rm GeV}\,\left(\frac{106.75}{\gs}\right)\,\eta\,.
\label{eq:N_degeneracy}
\end{equation}
As shown in figure~\ref{fig:deg23}, this amount of degeneracy is achievable in large region
of parameter space especially with the enhancement from $\eta$. 

For scenario $(i)$, since $M_3 > M_2 \gg M_1$, while the standard resonant leptogenesis 
is not possible, leptogenesis is naturally realized in the supersymmetric framework. 
In this case, we can realize soft leptogenesis where the mechanism only relies on the first generation 
of right-handed sneutrinos and anti-sneutrinos (superpartners of $N_1$) 
with the required mass splitting provided by soft supersymmetry 
breaking term~\cite{Grossman:2003jv,DAmbrosio:2003nfv,Fong:2011yx}.
In soft leptogenesis, due to additional temperature dependence of the CP parameter, 
$\epsilon(T') = \bar\epsilon\, f(T')$ where $\bar\epsilon$ and $f(T')$ denote respectively 
the temperature independent and dependent pieces, the analysis is more involved. 
Nevertheless, we can still parametrize the final baryon asymmetry as follows
\begin{equation}
Y_B = - \frac{8}{23} \bar{\epsilon}\, \xi\, \eta Y^0_{\tilde N_1},
\end{equation}
where the right-handed sneutrino abundance at $\kappa_1 = 1$ is given by
\begin{equation}
Y^0_{\tilde N_1} \equiv \frac{45\zeta(3)}{\pi^4 g_\star},
\end{equation}
and $\xi$ is obtained from solving the Boltzmann equations and encapsulates 
the temperature dependence of the CP parameter. Solving for $\xi$ goes beyond 
the scope of this work though we do not expect enhancement beyond 
the factor $\eta$ due to hotness. This enhancement $\eta > 1$ allows the following 
rescaling of the soft supersymmetry breaking parameters related to $\tilde N_1$: 
either a smaller trilinear $A$ term which controls the strength of CP violation $A \to A/\eta$
or a larger bilinear $B$ term which controls the mass splitting of right-handed sneutrinos $B \to B\, \eta$ 
(For more details, see refs.~\cite{Grossman:2003jv,DAmbrosio:2003nfv,Fong:2011yx}).

Now one can ask about the possible $B-L$ asymmetry generated at higher scale
prior to the decay of $N_1$. In general, this asymmetry would survive the washout from $N_1$ 
interactions. Nevertheless, if we assume that the asymmetry is generated from the decays of $N_{2,\,3}$, 
it will be relatively suppressed compared to the asymmetry generated from $N_1$ decays 
because $N_2$ and $N_3$ are in thermal equilibrium with the SM sectors. 
Essentially, the suppression comes from the washout from $N_{2,\,3}$ interactions 
$\sim 1/K_{2,\,3}$ and also from the smaller relative abundance of $N_{2,\,3}$ by $1/\kappa_1^3$. 

Finally, we would like to comment on the realization of ARS leptogenesis
where the RHNs are required to be at around GeV scale. 
This can be realized in the case of $T_d \gtrsim T_{\rm BBN} \sim {\rm MeV}$ 
discussed in section~\ref{sec:DMonly}. For low scale RHNs, the constraint 
not to equilibrate the temperature between the dark and the SM sectors before DM freeze-out 
\eqref{eq:scatt_constraint_hierarchy} can easily be fulfilled.
In this mechanism, leptogenesis proceeds through lepton-number-conserving 
oscillation of GeV scale RHNs at $10^{6}\,{\rm GeV} \gtrsim T > T_{\rm EWsp}$ GeV where lepton asymmetries 
are induced in the RHN flavors. Through scatterings \eqref{eq:Gamma_scatt_gen},
lepton asymmetries in the lepton doublets can be generated. With an active electroweak sphalerons which 
couple to the lepton doublets, a nonzero baryon number is also induced. 
Since this mechanism is dominated by lepton-number-conserving processes, 
at least one flavor of RHNs (denote $N_d$) should remain thermally decoupled from the 
SM thermal bath until $T < T_{\rm EWsp}$ such that we end up with $Y_{B-L} = - Y_{\Delta N_d} \neq 0$ 
where $Y_{\Delta N_d}$ denote the asymmetry resides in $N_d$. 
Ref.~\cite{Asaka:2017rdj} showed that in the ARS leptogenesis, similar to the standard leptogenesis, 
if the RHNs are weakly coupled, initial condition becomes relevant. In particular, in certain parameter space, 
they found enhancement of several orders of magnitude due to large initial abundance of RHNs. 
We expect our scenario with hot RHNs to provide further enhancement on top of their consideration. 
Finally, we want to highlight that since ARS leptogenesis depends on \emph{oscillations}
instead of decays of RHNs, it can work even with $n=2$, i.e. 
two RHNs. The reason is that $N_i$ only get into equilibrium 
close to their masses $M_i$ (or slightly earlier) and for $M_i \sim \mathcal{O}$(GeV), 
it becomes possible that the two sectors do not thermalize during the ARS leptogenesis 
at $T \gtrsim 100$~GeV.

\section{Dark Matter Relic Abundance}
\label{sec:DM}

In the present scenario of hot leptogenesis, DM behaves like a collisionless cold WIMP, with a relic abundance determined by its thermally averaged annihilation cross section $\langle\sigma v_X\rangle$ into RHNs and its initial temperature.
In our scenario, the interaction between the dark and the SM sectors are through the RHNs. 
If we make sure that the two sectors never achieve thermal equilibrium before the DM freeze-out~\eqref{eq:scatt_constraint_hierarchy}, then the two sectors can have two independent temperatures as
we will assume here.
In a model independent framework, the maximal annihilation cross section compatible with unitarity is given in eq.~\eqref{eq:maxsv}, in the case where $\xi=1$.
DM freezes out when its annihilation rate equals the Hubble expansion rate
\begin{equation}
\left.\frac{\Gamma}{H}\right|_\text{FO}\equiv\left.\frac{n_X\,\langle\sigma v_X\rangle}{H}\right|_\text{FO}=1\,.
\end{equation}
If the unitarity bound is saturated ($\xi=1$), the maximal freeze-out then takes place at
\begin{equation}\label{eq:xpfomax}
\xpfo_\text{max}\simeq 34.8+\log\left[\frac{\gx}{\sqrt{\gn+0.5\left(\frac{5}{\kfo}\right)^4}}\,\frac{\xpfo_\text{max}}{34.8}\,\frac{10^5~\text{GeV}}{\mx}\right]\,,
\end{equation}
where $\gn$ corresponds to the number of relativistic degrees of freedom of the $N_i$ at freeze-out, i.e. $\gn=2$, 4 and 6 for scenarios $(i)$, $(ii)$ and $(iii)$, respectively.
Note that for $\kfo\gtrsim 4$, the dark sector starts to dominate the energy density of the Universe, and therefore $\xpfo$ tends to become independent of $\kfo$.\\

If one neglects for the moment the decays of the RHNs $N_i$, and hence the entropy injection, the Boltzmann equation that keeps track of the evolution of the DM number density $n_X$ reads
\begin{equation}
\dot n_X+3\,H\,n_X=\langle\sigma v_X\rangle\left(n_X^2-{n_X}_\text{eq}^2\right)\,,
\end{equation}
and admits the following analytical solution~\cite{Kolb:1990vq}
\begin{equation}\label{eq:Y_beforedilution}
Y_0\,\mx\simeq 3.8\times\frac{\sqrt{\gn+\frac{\gs}{\kfo^4}}}{\gs}\kfo^3\frac{\xpfo}{M_\text{Pl}}\,\frac{1}{\langle\sigma v_X\rangle}\,,
\end{equation}
where $Y\equiv\frac{n_X}{s}$ corresponds to the DM number density normalized by the SM entropy density, and $\langle\sigma v_X\rangle$ depends on $\mx$, $\xpfo$ and $\xi$.
Notice that $Y$ is defined as a function of the SM entropy density because it can be related to the today DM relic abundance by means of 
\begin{equation}\label{eq:OmegaDM}
\Omega_X = \frac{\mx\,n_X}{\rho_c}=\frac{\mx\,s_0\,Y_0}{\rho_c}\simeq\left(2.742 \times 10^8~\text{GeV}^{-1} \, h^{-2}\right) \, \mx \, Y_0\,, 
\end{equation}
where $\rho_c$ is the critical SM energy density and the subindices `$0$' refer to the values nowadays.
In order to match the DM relic density as measured by the Planck satellite~\cite{Ade:2015xua} one needs $Y_0\,\mx\simeq 4\times10^{-10}$~GeV.
Let us note that in the case where $\kfo=1$ and $\xpfo\simeq 25$, eqs.~\eqref{eq:Y_beforedilution} and~\eqref{eq:OmegaDM} give rise to the standard WIMP result: $\langle\sigma v_X\rangle\simeq 3\times 10^{-26}$~cm$^3$/s~\cite{Steigman:2012nb}.

The decays of the $N_i$ inject energy in the visible sector, diluting both the DM and the lepton asymmetry.
The total dilution factor $D$
\begin{equation}
D=\left(\frac{\kfo}{\kt}\right)^3\left(1+ \frac{45 \zeta(3)}{\pi^4 \gs} \frac{\mn}{\Td} \kt^3 \right)^{3/4}
\label{eq:TOTALdilution}
\end{equation}
has two contributions: one due to the late decay of $N_1$ (eq.~\eqref{eq:dilution}), second bracket, and the other due to the decays of $N_2$ and $N_3$, first bracket, only present in scenarios $(ii)$ and~$(iii)$.
Let us remember that these dilution factors are defined as the SM entropy ratios after and before the decays.

The efficiency factor $\eta_\text{DM}$ in the DM production that quantifies the gain from hotness and the loss from dilution could be defined as $\eta_\text{DM}\equiv\kfo^3/D$.
However, one may note that this quantity reduces to the efficiency $\eta$ defined in eq.~\eqref{eq:efficiency}, $\eta_\text{DM}=\kfo^3/D=\kt^3/d=\eta$.

The final DM relic abundance, given by eqs.~\eqref{eq:Y_beforedilution} and~\eqref{eq:TOTALdilution}, depends on 5 free parameters: the DM and the lightest RHN mass ($\mx$ and $\mn$), $\xpfo$ and the ratio of temperatures $\kfo$ at the DM freeze-out, and the SM temperature $\Td$ at which $N_1$ decays.
These parameters are allowed to vary in the ranges:
\begin{eqnarray}
1 < & \kfo &\,,\label{eq:kfo}\\
3 \lesssim &\xpfo& \leq \xpfo_\text{max}\,,\\
&\mn& < \frac{\mx}{\xpfo}\,,\\
{\Td}_\text{min} < &\Td& < \frac{\mn}{\kt}\,.  \label{eq:Td} 
\end{eqnarray}
We focus on cases where, at the DM freeze-out, the dark sector is warmer than the SM and the DM is non-relativistic while $N_1$ is relativistic, eq.~\eqref{eq:N1_relativistic}.
The latter condition is important in the context of leptogenesis in order to avoid a strong Boltzmann suppression for $N_1$, and hence for the generation of the baryonic asymmetry.
Additionally, $\xpfo_\text{max}\equiv\xpfo_\text{max}(\mx,\,\kfo)$ is defined by the unitarity bound in eq.~\eqref{eq:xpfomax}.
Finally, we impose  
the condition such that $N_1$ does not thermalize both the dark and the SM sectors, eq.~\eqref{eq:outofeq}.
The lower bound on $\Td$, ${\Td}_\text{min}$, depends on the scenario considered either eq.~\eqref{eq:decaying_temperature_lower_bound} 
or eq.~\eqref{eq:reactivated_bound}.
The ranges defined in eqs.~\eqref{eq:kfo} to~\eqref{eq:Td} will be systematically used in all the following numerical analysis.

\subsection{Dark Matter Only}\label{sec:DMonly}

\begin{figure}[t!]
\centering
\includegraphics[width=0.47\textwidth]{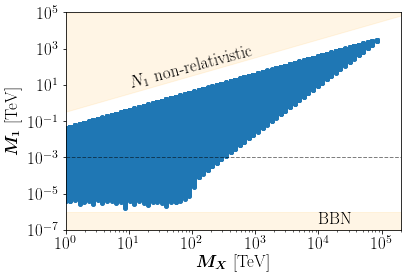}
\includegraphics[width=0.47\textwidth]{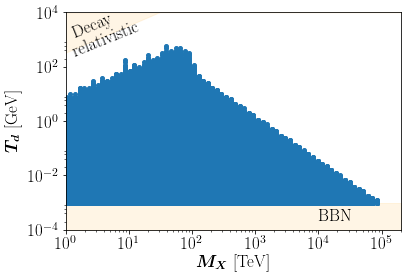}
\includegraphics[width=0.47\textwidth]{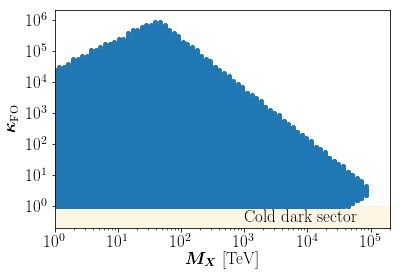}
\includegraphics[width=0.47\textwidth]{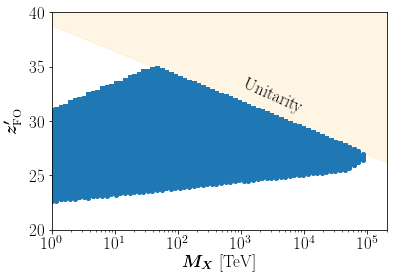}
\caption{Dark Matter only. Scan over the ranges defined in eqs.~\eqref{eq:kfo} to~\eqref{eq:Td}. The blue points reproduce the observed DM abundance.
The light orange bands are excluded because they violate the 
various bounds we impose. The dashed line in the top left plot indicates the regime favorable 
for ARS leptogenesis.
}
\label{fig:scanDM}
\end{figure}
Figure~\ref{fig:scanDM} presents the parameter space that gives rise to the observed DM relic abundance (blue regions).
The light orange bands are excluded because they violate the various bounds discussed earlier.
In the case where one only wants to reproduce the DM abundance, ${\Td}_\text{min}$ can go down to 1~MeV, in order to avoid Big Bang Nucleosynthesis (BBN) bounds~\cite{Cyburt:2015mya}.
In this case, DM can be as heavy as $\mathcal{O}(100)$~PeV without exceeding the measured cosmological DM density~\cite{Berlin:2016vnh, Berlin:2016gtr}, due to a large dilution factor produced by the late decay of a heavy $N_1$ ($\mn\simeq\mathcal{O}(1)$~PeV) and by an annihilation cross sections close to the unitarity bound (i.e. $\xpfo\simeq\xpfo_\text{max}$).
Let us note that at the moment of DM freeze-out, the dark sector can be much warmer than the visible sector ($\kfo\gg 1$), which again implies a highly populated dark sector and a potentially large dilution factor.
We emphasize that even if the dark sector is warmer than the SM, at freeze-out DM is always cold, with $\xpfo\simeq 25-30$.

\begin{figure}[t!]
\centering
\includegraphics[width=0.47\textwidth]{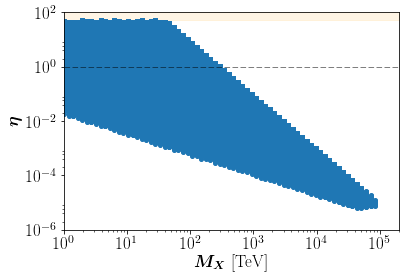}
\caption{Dark Matter only. Efficiency as a function of $\mx$. 
The efficiency is bounded from above by eq.~\eqref{eq:eta_max}.
}
\label{fig:effDM}
\end{figure}

The efficiency factor $\eta$ is shown in figure~\ref{fig:effDM}.
If $\eta\simeq 1$ the gain form hotness is compensated by the loss from dilution, and therefore the value for the annihilation cross section is close to the usual thermal one, $\langle\sigma v_X\rangle\simeq$~few~$\times 10^{-26}$~cm$^3$/s.
As expected from the unitarity bound, $\eta\simeq 1$ can be reached for $\mx\lesssim\mathcal{O}(100)$~TeV.
Lower efficiencies require lower cross sections, which allow the DM to be as heavy as $\mathcal{O}(100)$~PeV.
On the contrary, higher production efficiencies can happen with also higher annihilation cross section, which boost the indirect detection prospects (section~\ref{ID}). 
Let us remind that the efficiency is bounded from above by eq.~\eqref{eq:eta_max}.

Notice that in this scenario, the RHNs can be as heavy as PeV down to as light as MeV.
However, as we will see in the subsequent sections, for $M_1 \gtrsim 4$ TeV or $M_1 < T_{\rm EWsp}$, 
the decay temperature is too low for successful leptogenesis. Nevertheless, for $M_1 \sim $ GeV, 
ARS leptogenesis can be realized. We highlight this regime with a dashed line in the top left plot 
of figure~\ref{fig:scanDM}. In this regime, we can also have enhanced efficiency 
as illustrate in figure~\ref{fig:effDM}.

\subsection[Scenario $(i)$]{Scenario $\boldsymbol{(i)}$}

\begin{figure}[t!]
\centering
\includegraphics[width=0.47\textwidth]{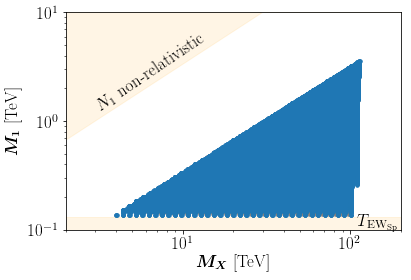}
\includegraphics[width=0.47\textwidth]{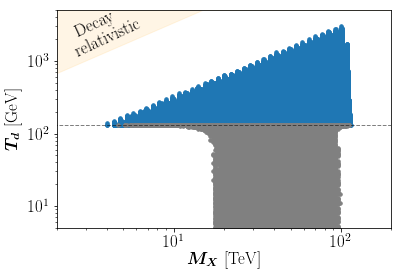}
\includegraphics[width=0.47\textwidth]{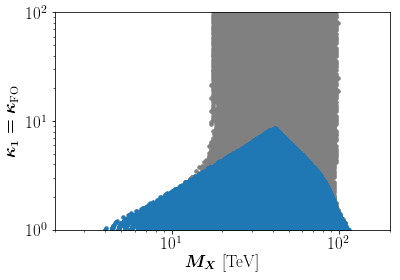}
\includegraphics[width=0.47\textwidth]{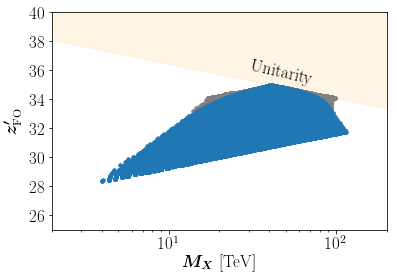}
\caption{Scenario $(i)$.
Scan over the ranges defined in eqs.~\eqref{eq:kfo} to~\eqref{eq:Td}. 
All the points reproduce the observed DM abundance, the neutrino masses and give rise to a successful leptogenesis.
Blue and gray regions correspond to $\Td$ greater and smaller that $T_\text{EWsp}$, respectively.
The light orange bands are excluded because they violate the various bounds we impose.
}
\label{fig:scan}
\end{figure}
\begin{figure}[t!]
\centering
\includegraphics[width=0.47\textwidth]{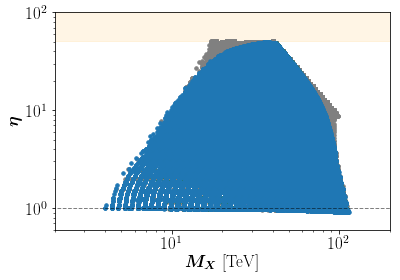}
\includegraphics[width=0.47\textwidth]{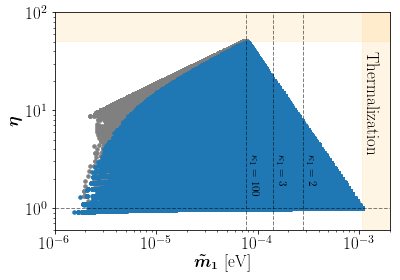}
\caption{Scenario $(i)$. Efficiency as a function of $\mx$ and $\tilde m_1$.
}
\label{fig:dil-eff}
\end{figure}

Scenario $(i)$ corresponds to the case where both $N_2$ and $N_3$ are very heavy ($m_{N_3}>m_{N_2}\gg T'_{RH}>\mx>\mn$) and therefore irrelevant for the DM phenomenology. 
In this case, $N_1$ will inherit the dark sector temperature and we have $\kt=\kfo$.
Figure~\ref{fig:scan} presents the parameter space that simultaneously generates the observed DM relic abundance and a successful leptogenesis.
The blue regions correspond to the standard case where $N_1$ decays before the electroweak sphalerons freeze-out, $\Td>T_\text{EWsp}$.
In this case, the dark sector temperature is bounded from above by eq.~\eqref{eq:kappa1_bound}.

Alternatively, the gray region corresponds to a less strict case, $\Td<T_\text{EWsp}$, where the SM thermal bath is reheated from $N_1$ decays and the electroweak sphalerons are `reactivated' to convert the lepton asymmetry to baryon asymmetry, section~\ref{sec:connection_DM}.
In this case, we demand $\tilde T$, the SM temperature after the decay of $N_1$ defined in eq.~\eqref{eq:Ttilde}, to be greater than $T_\text{EWsp}$ for the reactivation to take place 
and in principle $\kfo$ can take very large values, bounded eventually 
by $T_d > T_{\rm BBN}$. 

The typical allowed parameter spaces in the cases $\Td>T_\text{EWsp}$ and $\Td<T_\text{EWsp}$ are comparable, up to the fact that in the latter scenario $\kfo$ is allowed to reach higher values. 
While $M_X$ ranges from 4.5~TeV to 120~TeV, $M_1$ is bounded from $T_\text{EWsp}$ and $\simeq 4$~TeV. 
As we discussed in section \ref{sec:realizations_leptogenesis}, for this mass range of $N_1$, 
leptogenesis has to proceed with some amount of resonant enhancement. Since we
only have one light RHN with $M_1 \ll M_{2,\,3}$, resonant leptogenesis can proceed through soft leptogenesis.  
Figure~\ref{fig:dil-eff} shows the efficiency $\eta$ as a function of $\mx$ and $\tilde m_1$ 
(defined in eq.~\eqref{eq:meff}). 
In the right plot of figure~\ref{fig:dil-eff}, we show the maximum value of $\tilde m_1$ for various 
$\kappa_1$ beyond which the dark and the SM sectors will thermalize.
Since $\tilde m_1$ is bounded from below by the lightest light neutrino mass $m_l$,
eq.~\eqref{eq:ml_bound}, if $m_l$ is measured, the allowed range of $\eta$ is also 
constrained. Only if $m_l \leq 7.7 \times 10^{-5}$ eV, a maximal $\eta$ of 51.6 can be achieved.
For $m_l > 7.7 \times 10^{-5}$ eV, we have $\eta < 51.6$ while for $m_l > 1 \times 10^{-4}$~eV, 
thermalization between the dark and SM sectors occurs. If measurements fall outside the allowed
regime for e.g. $m_l = 6 \times 10^{-4}$ eV and $\eta = 20$, this will imply either our scenario 
does not hold or a scenario with more RHNs $n > 3$. 
Along the same line, if $m_l$ is measured to be greater than $7.7 \times 10^{-5}$ eV, 
we can also bound the largest $\kappa_1$ allowed for hot leptogenesis.

\subsection[Scenarios $(ii)$ and $(iii)$]{Scenarios $\boldsymbol{(ii)}$ and $\boldsymbol{(iii)}$}

\begin{figure}[t!]
\centering
\includegraphics[width=0.47\textwidth]{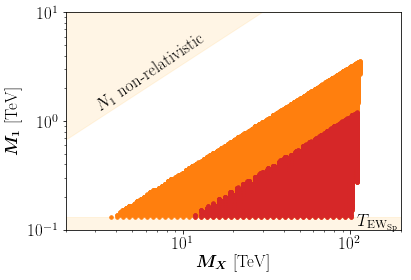}
\includegraphics[width=0.47\textwidth]{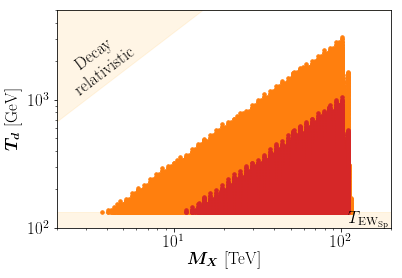}
\includegraphics[width=0.47\textwidth]{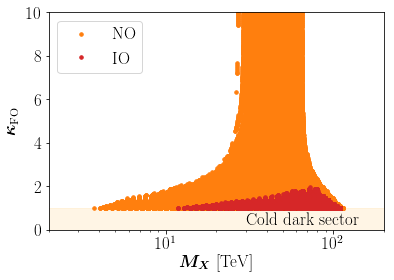}
\includegraphics[width=0.47\textwidth]{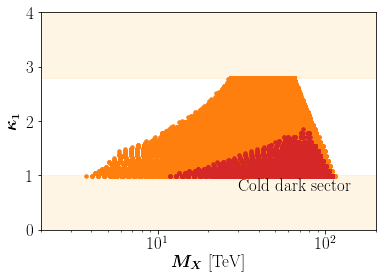}
\includegraphics[width=0.47\textwidth]{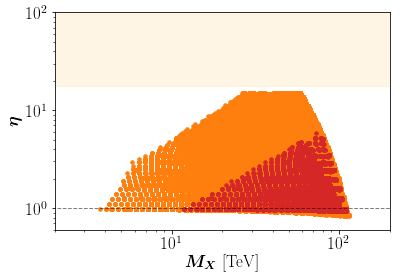}
\includegraphics[width=0.47\textwidth]{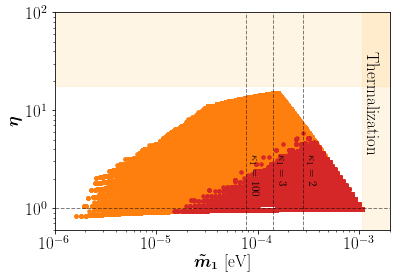}
\caption{Scenario $(ii)$.
All the points reproduce the observed DM abundance and are viable for hot leptogenesis.
The normal and inverse neutrino mass orderings are shown in orange and red, respectively.
The light orange bands are excluded because they violate the bounds we impose.
}
\label{fig:scan2}
\end{figure}
\begin{figure}[t!]
\centering
\includegraphics[width=0.47\textwidth]{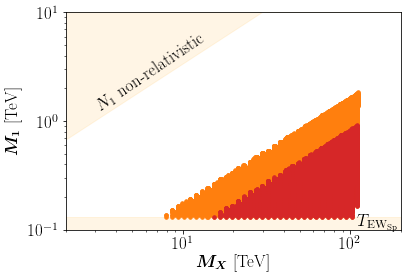}
\includegraphics[width=0.47\textwidth]{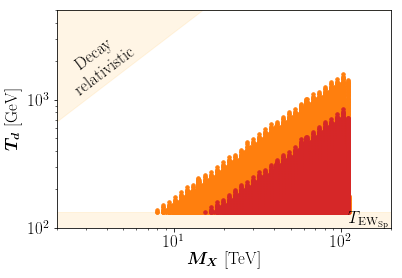}
\includegraphics[width=0.47\textwidth]{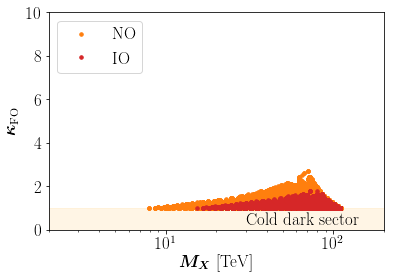}
\includegraphics[width=0.47\textwidth]{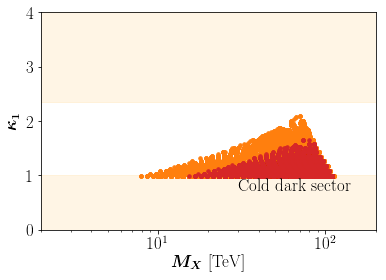}
\includegraphics[width=0.47\textwidth]{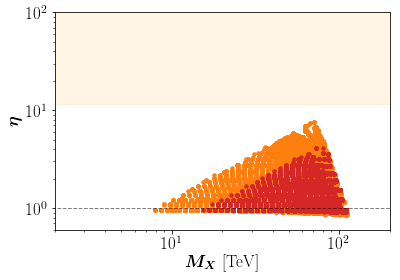}
\includegraphics[width=0.47\textwidth]{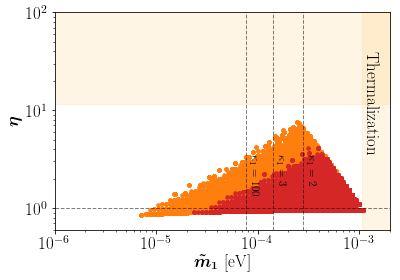}
\caption{Scenario $(iii)$.
All the points reproduce the observed DM abundance and are viable for hot leptogenesis.
The normal and inverse neutrino mass orderings are shown in orange and red, respectively.
The light orange bands are excluded because they violate the bounds we impose.
}
\label{fig:scan3}
\end{figure}
\begin{figure}[t!]
\centering
\includegraphics[width=0.47\textwidth]{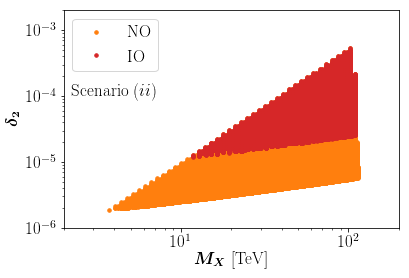}
\includegraphics[width=0.47\textwidth]{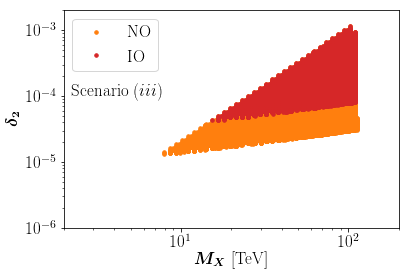}
\caption{The mass splitting defined in eq.~\eqref{eq:delta_j} for scenarios $(ii)$ and $(iii)$ 
consistent with the decoherence conditions eqs.~\eqref{eq:decoh_a} and \eqref{eq:decoh_b}. 
}
\label{fig:deg23}
\end{figure}
In scenario $(ii)$, in addition to $N_1$, we have $N_2$ which is also lighter than the DM while in scenario $(iii)$, 
$N_3$ is also lighter than the DM.
In order to guarantee that hot leptogenesis can proceed, one has to ensure that the scatterings of the RHN with the SM particles do not reach equilibrium before the DM freeze out, eq.~\eqref{eq:scatt_constraint_hierarchy}.
Figures~\ref{fig:scan2} and~\ref{fig:scan3} depict the parameter space compatible with the DM relic abundance and hot leptogenesis scenarios $(ii)$ and $(iii)$, respectively.
The normal and inverse neutrino mass orderings are shown in orange and red respectively.

Furthermore, in order to prevent the thermalization of $N_1$ with that of the SM, 
we have to make sure that oscillations of the RHNs are not effective 
due to decoherence as discussed in section~\ref{sec:coh}. The two decoherence conditions, 
eqs.~\eqref{eq:decoh_a} and~\eqref{eq:decoh_b}, depend on the mass splitting of the RHNs~\eqref{eq:delta_j}.
In figure~\ref{fig:deg23}, we present the mass splitting between the two lightest RHNs~$\delta_2$ as defined 
in eq.~\eqref{eq:delta_j} consistent with the decoherence conditions. 
From the plot, we see that successful resonant leptogenesis consistent with eq.~\eqref{eq:N_degeneracy} 
is possible.

\section{Indirect Detection}
\label{ID}

\begin{figure}[t!]
\centering
\includegraphics[width=0.47\textwidth]{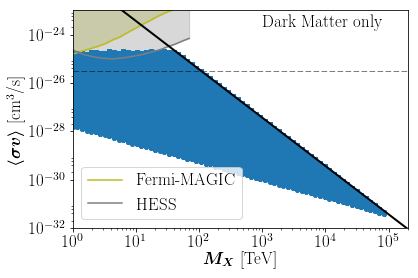}
\caption{Dark Matter only.
Upper limits on the thermally-averaged annihilation cross section, assuming an annihilation into $b\bar b$ pairs via an $s$-channel.
The blue and orange lines correspond to the limits obtained by combining observations with MAGIC and Fermi-LAT, and with H.E.S.S., respectively.
The thick black line represents the upper bound on cross section due to unitarity.
}
\label{fig:IDDM}
\end{figure}
\begin{figure}[t!]
\centering
\includegraphics[width=0.47\textwidth]{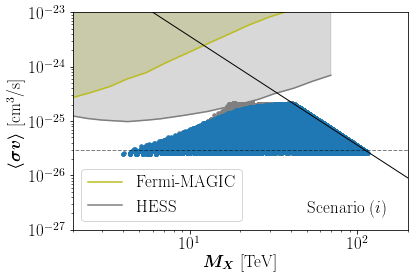}\\
\includegraphics[width=0.47\textwidth]{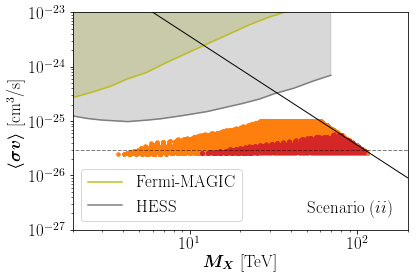}
\includegraphics[width=0.47\textwidth]{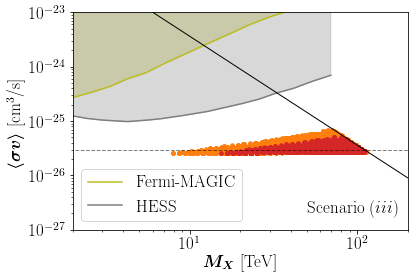}
\caption{Scenarios $(i)$, $(ii)$ and $(iii)$.
Upper limits on the thermally-averaged annihilation cross section, assuming an annihilation into $b\bar b$ pairs via an $s$-channel.
The blue and orange lines correspond to the limits obtained by combining observations with MAGIC and Fermi-LAT, and with H.E.S.S., respectively.
The thick black line represents the upper bound on cross section due to unitarity.
}
\label{fig:ID}
\end{figure}
DM can annihilate into the RHNs, which subsequently decay into $h\,\nu$, $Z\,\nu$ and $W^\pm\,l^\mp$.
One can look for stable SM particles issued from these processes, and in particular for gamma-rays.
Figure~\ref{fig:IDDM} shows the $95\%$ CL upper limits on the thermally-averaged cross section for DM particles annihilating into $b\bar b$ pairs.
The upper band corresponds to the limit obtained by combining 158 hours of Segue 1 observations with MAGIC, with 6-year observations of 15 dwarf satellite galaxies by the Fermi-LAT~\cite{Ahnen:2016qkx}.
The lower band corresponds to the limit using 10 years of galactic center observations by H.E.S.S. array of ground-based Cherenkov telescopes~\cite{Abdallah:2016ygi}, assuming an Einasto profile.
The black thick line is to the upper bound on the thermally averaged cross section, eq.~\eqref{eq:maxsv}, assuming an $s$-wave annihilation.
Figure~\ref{fig:IDDM} also overlays, in blue, the region where the observed DM relic abundance can be reproduced, irrespectively of the leptogenesis considerations.
For DM masses $\mx\lesssim 100$~GeV the cross section $\langle\sigma v\rangle$ can take values of the order of the usual thermal one $\simeq 3\times 10^{-26}$~cm$^3$/s~\cite{Steigman:2012nb} (dashed horizontal line), which corresponds to $\eta\simeq 1$.
If the gain from hotness is not compensated by the dilution ($\eta>1$), a large cross section is needed, in order to reproduce the observed DM abundance.
This scenario is particularly interesting because it falls in the ballpark tested by actual and future detectors.
In fact, a small portion of the parameter space is already in tension with observations from H.E.S.S.; however there are large inherent astrophysical uncertainties associated with the central region of the Milky way~\cite{Zemp:2008gw, Pato:2010yq, Bernal:2014mmt, Bernal:2015oyn, Bernal:2016guq, Benito:2016kyp}.
Future ground-based imaging air Cherenkov telescopes like CTA~\cite{Doro:2012xx} and HAWK~\cite{Abeysekara:2014ffg} will increase the sensitivity in this region of the parameter space.
Let us note that ref.~\cite{Batell:2017rol} also discussed the indirect detection prospects of a similar model, but focusing on lighter DM, in the sub-TeV ballpark.
Additionally, let us also note that a more refined analysis of the exclusion bounds could be done, as the one in ref.~\cite{Campos:2017odj}, studying the precise final states ($h\,\nu$, $Z\,\nu$ and $W^\pm\,l^\mp$) and not the simplistic $b\bar b$, however it is out of the scope of this paper.

For scenarios $(i)$, $(ii)$ and $(iii)$, similar results are presented in figure~\ref{fig:ID}.
Notice that due to the upper bound on $\eta$ as shown in eq.~\eqref{eq:eta_max}, 
the maximum possible enhancement to the cross section is bounded by $\max (\eta)$ times 
the thermal cross section. This also represents a prediction of our model.

Alternatively, on can look for a continuum neutrino signal from DM annihilations in the Milky Way.
Both the IceCube~\cite{Aartsen:2014hva, Aartsen:2016pfc, Aartsen:2017ulx} and the ANTARES~\cite{Adrian-Martinez:2015wey, Albert:2016emp} collaborations have looked for this signals, founding no significant excess of neutrinos over the background of neutrinos produced in atmospheric air showers from cosmic ray interactions.
If the RHNs are sufficiently long-lived, one can also look for neutrino lines and other spectral features induced by its decay~\cite{Garcia-Cely:2016pse, Queiroz:2016zwd, ElAisati:2017ppn}.
These channels could be definitely used, however the limits do not constrain the present scenario.

Before closing this section, let us comment on other detection techniques.
On the one hand, DM direct detection is very suppressed because $(i)$ DM is in the multi-TeV ballpark and the maximal sensitivity of typical direct detection experiments is around 10 to 100~GeV, and $(ii)$ DM only directly couples to the RHN, so the elastic scattering interaction rates are loop-suppressed.
On the other hand, the DM production at colliders is again challenging because of the DM mass.
Additionally, this mechanism strongly depends on the production of the RHN, which in turn depends on the typically suppressed mixing with active neutrinos.

\section{Conclusions}
\label{sec:conclu}

In this work, we have studied the scenario where the dark sector has a higher temperature 
than the SM sector after inflation. We considered the scenario where thermal DM 
abundance is fixed through the freeze-out of their annihilations to RHNs. After the freeze-out, 
the abundances of RHNs for leptogenesis are also fixed. While the abundances of DM
and RHNs are enhanced by hotness, they are also subject to the dilution from hot RHNs late decays. 
We have showed that generically the efficiency $\eta$, which is defined as gain from hotness 
divided by the loss from dilution, is greater than 1 but limited from above at 51.6.

In our model, leptogenesis from decays of RHNs can proceed with an enhanced efficiency.
Nevertheless, resonant enhancement in CP violation is still required. 
We can summarize the reason as follows.
The unitarity bound on the thermal DM mass goes as 
$M_X \lesssim 10^5\,{\rm GeV}/\eta^{1/3}$.\footnote{The power of 
$\eta$ is not one as naively expected since the Hubble rate is also modified. 
The dependence is in general more complicated but for large $\eta$, the power goes as $1/3$.} 
Additionally, the Davidson-Ibarra bound for hierarchical RHNs in type-I leptogenesis 
goes as $M_1 \gtrsim 10^8\,{\rm GeV}/\eta$. Assuming $M_X \gg M_1$, the bounds above cannot be reconciled 
unless we have a very large $\eta \gtrsim 10^5$ which however cannot be achieved by the 
bound $\eta < 51.6$ obtained in this work. 
We concluded that soft leptogenesis can be realized in scenario $(i)$ while 
the standard resonant leptogenesis can be realized in scenarios $(ii)$ and $(iii)$.
Finally we also highlighted the possibility of realizing ARS leptogenesis 
with an enhanced efficiency.

For the thermal DM, we can have an enhancement in the DM annihilation cross section. 
Detecting indirect signatures from DM annihilations 
consistent with $XX \to NN$ ($N \to h \nu, Z \nu, W^{\pm} \ell^{\mp}$) 
with larger cross section than the thermal one $\sim 3 \times 10^{-26}\,{\rm cm}^3{\rm /s}$
will support our scenario. The enhancement from the thermal cross section up to 51.6 
represents a prediction of our model.

\section*{Acknowledgments}
We thank Enrico Nardi for careful reading and valuable comments on the manuscript.
We acknowledge Dan Hooper and Jacobo López-Pavón for discussions and comments.
NB is partially supported by the National Science Centre (Poland) research project, decision DEC-2014/15/B/ST2/00108 and by the Spanish MINECO under Grant FPA2014-54459-P.
CSF is supported by the São Paulo Research Foundation (FAPESP) under grants 2012/10995-7, 2013/13689-7 \& 2017/02747-7.
NB and CSF would like to acknowledge the hospitality of the University of Warsaw and Fermilab, respectively, while this work was being completed.
This project has received funding from the European Union's Horizon 2020 research and innovation programme under the Marie Skłodowska-Curie grant agreements 674896 and 690575; and from Universidad Antonio Nariño grant 2017239.
This manuscript has been authored by Fermi Research Alliance, LLC under Contract No. DE-AC02-07CH11359 with the U.S. Department of Energy, Office of Science, Office of High Energy Physics.

\bibliographystyle{JHEP}
\tiny
\bibliography{biblio}

\end{document}